# Stochastic convergence in per capita $CO_2$ emissions. An approach from nonlinear stationarity analysis[1]

María José Presno (mpresno@uniovi.es)
Manuel Landajo (landajo@uniovi.es)
Paula Fernández González (pfgonzal@uniovi.es)
Departament of Applied Economics, University of Oviedo, Avenida del Cristo s/n°, 33009 Oviedo, Asturias, Spain

ABSTRACT

This paper studies stochastic convergence of per capita $CO_2$ emissions in 28 OECD countries for the 1901-2009 period. The analysis is carried out at two aggregation levels, first for the whole set of countries (*joint analysis*) and then separately for developed and developing states (*group analysis*). A powerful time series methodology, adapted to a nonlinear framework that allows for quadratic trends with possibly smooth transitions between regimes, is applied. This approach provides more robust conclusions in convergence path analysis, enabling (a) robust detection of the presence, and if so, the number of changes in the level and/or slope of the trend of the series, (b) inferences on stationarity of relative per capita $CO_2$ emissions, conditionally on the presence of breaks and smooth transitions between regimes, and (c) estimation of change locations in the convergence paths. Finally, as stochastic convergence is attained when both stationarity around a trend and $\beta$-convergence hold, the linear approach proposed by Tomljanovich and Vogelsang (2002) is extended in order to allow for more general quadratic models. Overall, joint analysis finds some evidence of stochastic convergence in per capita $CO_2$ emissions. Some dispersion in terms of $\beta$-convergence is detected by group analysis, particularly among developed countries. This is in accordance with per capita GDP not being the sole determinant of convergence in emissions, with factors like search for more efficient technologies, fossil fuel substitution, innovation, and possibly outsources of industries, also having a crucial role.

[1] This manuscript is an early version of Presno, M.J., Landajo, M., and P. Fernández González (2018): "Stochastic convergence in per capita $CO_2$ emissions. An approach from nonlinear stationarity analysis," Energy Economics, Vol. 70, pp. 563-58.

## 1. INTRODUCTION

Scientists have grown conscious that emissions of greenhouse gases (GHGs) are major contributors to the global climate change and greenhouse effect. The fifth International Panel on Climate Change (IPCC) Report states "it is extremely likely [95 percent confidence] that more than half of the observed increase in global average surface temperature from 1951 to 2010 was caused by the anthropogenic increase in greenhouse gas concentrations and other anthropogenic forces together". Scientists were only 90% confident in 2007, and it was not until the IPCC's Second Assessment Report in 1995 that a 'discernible' human influence on global climate was identified.

Political and economic agents are now aware of the importance of taking measures to save energy and reduce GHG emissions to the atmosphere. Those measures -which seek to gain competitiveness and to achieve sustainable growth- are sponsored by international agreements and institutions such as the United Nations (United Nations, 2009), the International Energy Agency (IEA, 2011), Kyoto Protocol (1997), the Decision 406/2009/EC of the European Parliament to reduce GHG emissions by 20% by 2020 in relation to 1990 levels, the United Nations Climate Change Conferences in Doha (2012) and Warsaw (2013), and more recently the European Council (2014), that agreed on the 2030 Climate and Energy Policy Framework for the European Union[2].

To achieve success in environmental policy, convergence has become an issue under discussion, with the interest focussed on convergence of per capita emissions, as it may represent a more acceptable basis than absolute emission levels for political compromises (*e.g.*, Stegman, 2005; Aldy, 2006). Being as carbon dioxide is the most important anthropogenic GHG, in this paper we will focus on convergence in per capita $CO_2$ emissions. More precisely, we shall examine convergence in relative per capita $CO_2$ emissions, which entails that countries are moving towards a common standard of environmental performance, instead of following independent paths in pollution control (Lee and Chang, 2008).

The seminal paper on convergence in $CO_2$ emissions is due to Strazicich and List (2003), who test for stochastic and conditional convergence in 21 OECD countries. They apply panel unit root testing, finding significant evidence of stochastic convergence. Since then, some papers have appeared, considering various countries, periods, and statistical tools. Thus, Lanne and Liski (2004), using unit root tests allowing for structural breaks, study 15 developed countries and conclude that most of their series are non-stationary. Lee *et al.* (2008) also tackle the problem from the structural break unit root approach and conclude stochastic convergence. McKibbin and Stegman (2005) examine a range of variables related to climate change projections, with focus on per capita carbon emissions from fossil fuel. They conclude that no convergence is detected when a large cross section of countries is considered. When the analysis is restricted to OECD countries, the same paper finds some tendency towards convergence, although absolute convergence seems unlikely due to differences in fossil fuel endowments. Aldy (2006) applies unit root testing to a large sample of 88 countries, finding divergence, although the series appear to converge when the analysis is carried out for 23 OECD states.

[2] The European Council endorsed 4 targets: a binding EU target of 40% less GHG by 2030, compared to 1990; a target of at least 27% renewable energy consumption; a 27% energy efficiency increase; and the completion of the internal energy market by achieving the existing electricity interconnection target of 10% and linking the energy islands (Baltic states and the Iberian Peninsula).

From a panel unit root perspective, Westerlund and Basher (2008) –following the notion of convergence proposed by Evans (1998)- test for convergence in per capita carbon dioxide emissions on a collection of developed and developing countries. Romero-Ávila[3] (2008) and Lee and Chang (2009) extend the analysis by using the panel stationarity test of Carrión-i-Silvestre *et al.* (2005), which allows for the presence of structural breaks. Their results support the hypothesis of convergence. However, Barassi *et al.* (2008) apply a battery of panel stationarity and unit root tests, concluding that emissions have not converged among OECD countries during the study period. Lee and Chang (2008) examine stochastic convergence from the panel unit root perspective, also extending their analysis to $\beta$-convergence.

Nourry (2009) analyses stochastic convergence for $CO_2$ and $SO_2$ emissions, by applying a pairwise approach that considers all possible pairs of log per-capita emission gaps across economies. Their results reject the hypothesis of stochastic convergence in per capita emissions for both pollutants, even in the OECD sub-dataset.

Panopoulou and Pantelidis (2009) examine the evidence in favour of club convergence, finding that two convergence clubs -converging to different steady states- are observed in recent years: the one includes countries with high per capita $CO_2$ emissions whereas the other incorporates states having low per capita $CO_2$ emissions. Camarero *et al.* (2013;

[3] Romero-Ávila (2008) examines both stochastic and deterministic convergence.

2014) focus on convergence in eco-efficiency[4] and find different convergence clubs depending on the specific pollutant.

Nguyen-Van (2005) and Ezcurra (2007) apply nonparametric methods. The former finds little evidence of convergence for the whole dataset, while Ezcurra (2007) reports a convergence process along the study period, although he warns that it will not continue indefinitely. Jobert *et al.* (2010) exclusively deal with European countries, using a Bayesian shrinkage estimation method, to support the hypothesis of absolute convergence, whereas Ordás Criado and Grether (2011) base their analysis on the evolution of the spatial distributions over time.

Camarero *et al.* (2011) and Yavuz and Yilanci (2013) focus on the nonlinear aspects of convergence. The former applies a unit root test within a STAR framework, concluding that there exists no convergence among the 22 OECD countries analysed. Yavuz and Yilanci (2013), after rejecting the null of linearity, address convergence by using a TAR panel unit root test, concluding that the per capita $CO_2$ emissions of G7 countries diverged only when fossil fuel became the main source of productivity or in the event of an oil crisis.

In this paper we contribute to the debate by examining convergence of per capita $CO_2$ emissions within a nonlinear framework. Concretely, we consider a quadratic trend model with smooth transition between regimes. Results from new time series methodologies providing more robust conclusions about the convergence paths are also incorporated.

First, we apply stationarity tests around a quadratic trend allowing for breaks and smooth (*i.e.*, gradual) changes. Classical stationarity analysis has focused on unit root testing – where the null hypothesis of a unit root is tested against the alternative of trend stationarity. In our study we formulate the testing problem in the opposite direction: the null of trend stationarity (around a quadratic trend with breaks or smooth transitions) will be tested against a unit root alternative. This is in line with reasoning by Camarero *et al.* (2011), where it is pointed out that the first source of $CO_2$ emissions is economic activity -which exhibits a cyclical behaviour- so a similar functional form is also likely in $CO_2$ (Lee and Chang, 2009; Lanne and Liski, 2004). Additionally, most previous papers have relied on quadratic trends –usually assuming an inverted U functional form[5]- in order to model the long-run relationship between emissions and income. Finally, in economic activity, transition between regimes occurs gradually –as it is characterised by a delay between shocks and the reaction of economic agents-, so $CO_2$ emissions are expected to exhibit the same behaviour. With the remarkable exceptions of Camarero *et al.* (2001) and Yavuz and Yilanci (2013), no other analysis has focused on the nonlinear patterns of convergence in $CO_2$ emissions.

[4] This concept refers to the ability to produce more goods and services with a lesser impact on the environment and lesser consumption of natural resources.

[5] Since the seminal paper of Grossman and Krueger (1995) numerous studies have focused on the relationship between income and $CO_2$ emissions, postulating the existence of an inverted U relationship (the Environmental Kuznets Curve) between economic development and environmental degradation.

Secondly, we shall apply a robust methodology for the treatment of breaks and smooth transitions in stationarity testing. Perron and Yabu (2009), Harvey *et al.* (2010) and Kejriwal and Perron (2010) have proposed approaches that are robust to both unit root and stationary errors, in order to be able to test for stability of the trend function and to obtain consistent estimates for the true number of breaks in the series. In their original versions these techniques considered linear trends. Here we exploit Presno *et al.* (2014) extension in order to allow for quadratic trends. Once the number of changes has been assessed, we apply two alternative types of stationarity tests in order to clear up the stochastic properties of the data. The first test allows for the presence of breaks –*i.e.*, instant changes- whereas the second one includes smooth transitions instead.

Third, knowing the stochastic characteristics of the time series is useful in order to tackle consistent estimation of the change dates (respectively, breaks and midpoints in smooth transitions). Concretely, when changes in level and/or slope are detected, consistent estimates for the break dates are obtained from either a level or a first-differenced specification, according to whether a stationary process is present or not.

Finally, since stochastic convergence is attained when both stationarity around a trend and $\beta$-convergence are verified, we adapt the mainstream $\beta$-convergence methodology to the quadratic case with the aim of completing the analysis.

Summing up, in order to analyse –within a nonlinear framework- stochastic convergence of per capita $CO_2$ emissions in 28 OECD countries during the 1901-2009 period, we apply a time series methodology that enables (i) robust detection of the presence and number of changes in the trend of relative per capita $CO_2$ emissions, (ii) inferences on stationarity of the series in the presence of breaks and smooth transitions, (iii) estimation of change date locations in the convergence paths, and (iv) analysis of $\beta$-convergence.

The rest of the paper is structured as follows: Section 2 introduces the methodology. Section 3 reports the empirical results with a discussion. The paper closes with a summary of conclusions.

## 2. METHODOLOGICAL ISSUES

Convergence analysis from a time series perspective rests on the concept of stationarity, that is, the idea that shocks only have a temporary effect. Adapting the methodology proposed by Carlino and Mills (1993), our approach relies on testing the stochastic properties of the logarithm of the ratio of per capita $CO_2$ emissions in each country to average per capita emissions for the whole sample of countries. In this section we outline the econometric methodology for the analysis in the nonlinear (quadratic) case. Since stationarity testing depends on the presence and number of structural changes in the series, that number must be estimated in a previous stage by relying on a suitable approach that is robust to stationary/integrated errors. In a second stage the stochastic properties of the data are examined by testing for the null of stationarity around a quadratic trend with changes, allowing for two possibilities (respectively, breaks and smooth transitions). Knowledge about the properties of the series allows us to estimate change locations. Finally, $\beta$-convergence is examined.

### 2.1. Estimation of the number of changes

A well-known circularity problem exists between tests on the parameters of the trend function and unit root/stationarity testing. On one hand, sufficient knowledge of the properties of the series is necessary in order to test for structural breaks[6]. On the other hand, information about the number of changes is vital in order to devise unit root and stationarity tests with good properties and to avoid power loss. Perron and Yabu (2009), Kejriwal and Perron (2010) and Harvey *et al.* (2010) elegantly solved the circular problem by proposing a methodology that allows to test for the presence of structural changes in the linear trend function of a time series without any prior knowledge about the characteristics of the noise component. More recently, Presno *et al.* (2014) extended these approaches in order to incorporate a quadratic trend to cope with nonlinearities. Now we introduce these methodologies.

***Test for the presence of a structural change***

[6] Implementation of the test for structural breaks in the level of the series entails different limiting distributions depending on whether a unit root is present or not, and inference from first-differenced data - which conveys to assume a unit root- leads to tests with poor properties when the series contains a stationary component.

Perron and Yabu (2009) propose an approach to test for stability of the trend function, assuming the following data generating process:

$$\begin{aligned} y_t &= x_t^{'}\Psi + u_t \\ u_t &= \alpha u_{t-1} + \nu_t \\ \nu_t &= d(L)e_t \end{aligned} \qquad (1)$$

for $t$=1,…,$T$; where $x_t$ and $\Psi$ are, respectively, an ($r$ $x$ 1) vector of deterministic components and a vector of unknown parameters; $d(L)=\sum_{i=0}^{\infty} d_i L^i$ , with $\sum_{i=0}^{\infty} i|d_i| < \infty$ and $d(1)\neq 0$; $e_t \approx i.i.d.(0,\sigma^2)$, and $u_0$ is a constant. Assuming that -1<$\alpha$≤1, the case $\alpha$=1 corresponds to a difference stationary (or $I$(1)) process, whereas the series is trend stationary ($I$(0)) for -1<$\alpha$<1, in both cases with a possibly broken trend.

Perron and Yabu (2009) examine two cases: a shift in either intercept or slope (their Models I and II, respectively) and a shift in both intercept and slope (Model III) at instant $T_b=[\lambda T]$ for $\lambda\in(0,1)$, where [.] denotes the largest integer being less than or equal to the argument. Presno *et al.* (2014) adapt the procedure to the quadratic trend case, focusing on Model III: $x_t=(1,t,t^2,DU_t,DT_t)'$, $DU_t=1(t>T_1)$, $DT_t=1(t>T_1)(t-T_1)$, with 1(.) being the indicator function, and $\Psi=(\beta_0,\beta_1,\beta_2,\delta_1,\eta_1)'$. Under this model, two cases (respectively, the "general" and the "unrestricted" one) may be considered. For the "general" case, the hypothesis of interest is $\delta_1=\eta_1=0$. The "unrestricted" case[7] tests for stability of the slope parameter, allowing the intercept to vary between regimes, so the hypothesis of interest is $\eta_1=0$, and the same critical values corresponding to Model II are used. The test statistic is the *Exp* functional of the Wald test (*ExpW*).

### *Sequential test for the number of breaks*

Subsequently, Kejriwal and Perron (2010) proposed a sequential procedure that enables consistent estimation of the number of breaks without prior knowledge on the nature of persistence in the noise component. The procedure is as follows: conditionally on rejection in Perron and Yabu (2009) test, the break date is estimated and then the methodology of Perron and Yabu is newly applied in order to test for the presence of an

[7] In order to distinguish between changes in level and slope, Kejriwal and Lopez (2013) recommend implementing the "unrestricted" proposal, as they show that the test for Model III also has power against processes characterized by shifts only in level.

additional break in each subsample. The test statistic for the null of one versus two breaks is:

$$ExpW(2/1) = \max_{1 \le i \le 2} \left\{ ExpW^{(i)} \right\} \tag{2}$$

where $ExpW^{(i)}$ is the one-break test in sample *i*.

***Test for breaks in level***

The above procedure is not adapted to investigate the presence of breaks that affect solely the level of the series. So, for the linear case -and conditionally on a stable underlying slope- Harvey *et al.* (2010) proposed robust tests to detect multiple changes in level. Presno *et al.* (2014) extended this methodology to the quadratic case, considering the following model:

$$\begin{aligned} y_t &= \beta_0 + \beta_1 t + \beta_2 t^2 + \sum_{i=1}^{n} \delta_i DU_t([\lambda_i T]) + u_t, t = 1,..., T \\ u_t &= \rho u_{t-1} + \varepsilon_t, \; t = 2,..., T \end{aligned} \tag{3}$$

where $\lambda_i \in \Lambda$, $\Lambda = [\lambda_L, \lambda_U]$, with $\lambda_L$ and $\lambda_U$ being trimming parameters satisfying $0 < \lambda_L < \lambda_U < 1$. The null $\delta_i = 0$ (*i*=1,…, *n*) is tested against the alternative that there is at least one break in level.

The test statistics are:

$$\begin{aligned} M &= \max_{t \in \Lambda} \left| M_{t,[mT]} - \hat{\beta}_1 \left[ \frac{m}{2} T \right] - \hat{\beta}_2 (2t+1) \left[ \frac{m}{2} T \right] \right| \\ S_0 &= (\hat{\omega}_\upsilon)^{-1} T^{-1/2} M \\ S_1 &= (\hat{\omega}_u)^{-1} T^{1/2} M \end{aligned} \tag{4}$$

with

$$M_{t,[mT]} = \frac{\sum_{i=1}^{\left[\frac{m}{2}T\right]} y_{t+i} - \sum_{i=1}^{\left[\frac{m}{2}T\right]} y_{t-i+1}}{\left[\frac{m}{2}T\right]}$$

for a window width $m$ satisfying $n \leq 1 + \left[ \frac{\lambda_U - \lambda_L}{m} \right] = n_{\max}$. $\hat{\beta}_1$ and $\hat{\beta}_2$ are OLS estimators for the linear and the quadratic term coefficients, respectively, and $\hat{\omega}_\upsilon$ and $\hat{\omega}_u$ are long-run variance estimates for the case of $I(1)$ and $I(0)$ shocks.

The test statistic is:

$$U = \max \left\{ S_1, \left( \frac{cv_\xi^1}{cv_\xi^0} \right) S_0 \right\} \tag{5}$$

where $cv_\xi^1$ and $cv_\xi^0$ are critical values for $S_1$ and $S_0$ at significance level $\xi$. The null is rejected if $U > \kappa_\xi cv_\xi^1$, for a positive scaling constant $\kappa_\xi$, and this would inform us that at least one level break is present. Harvey *et al.* (2010) also propose a sequential procedure in order to determine the number of level breaks ($n_U$).

## 2.2. Stationarity testing under nonlinear trends

Once the number of changes in the time series has been estimated, stationarity analysis is carried out through the proposal of Landajo and Presno (2010), which extends classical KPSS testing to nonlinear models with endogenously determined changes –including abrupt changes and smooth transitions between regimes.

The error-components model is as follows:

$$y_{t,T} = \mu_t + f(t/T, \boldsymbol{\theta}) + \varepsilon_t, \tag{6}$$

$$\mu_t = \mu_{t-1} + u_t; \;\; t = 1,...,T; \;\; T = 1,2,...$$

where $\{\varepsilon_t\}$ and $\{u_t\}$ are independent zero-mean error processes with respective variances $E(\varepsilon_t^2) = \sigma_\varepsilon^2 > 0$ and $E(u_t^2) = \sigma_u^2 \geq 0$; $\{\mu_t\}$ starts with $\mu_0$, which is assumed to be zero. $f(t/T, \boldsymbol{\theta})$ is a smooth function of time with $\boldsymbol{\theta}$ being a vector of free parameters. In order to allow for smooth transitions, we consider logistic sigmoid changes under the following specifications:

Model I:

$$f(t/T, \boldsymbol{\theta}) = \beta_0 + \beta_1 t/T + \beta_2 (t/T)^2 + \sum_{k=1}^{n} \delta_k \left[1 + \exp\{-\gamma_k (t/T - \lambda_k)\}\right]^{-1} \tag{7}$$

and Model III:

$$f\left(t/T,\boldsymbol{\theta}\right)=\beta_0+\beta_1 t/T+\beta_2\left(t/T\right)^2+\sum_{k=1}^{n}\left(\delta_k+\eta_k t/T\right)\left[1+\exp\left\{-\gamma_k\left(t/T-\lambda_k\right)\right\}\right]^{-1} \quad (8)$$

where $\lambda_k \in [0,1]$ determines the relative position of the timing of the transition midpoint $T_{b,k}$ into the sample and $\gamma_k$ controls the speed of transition (gradual for small $\gamma_k$, and approaching a break as $\gamma_k$ increases). Model I enables analysis of series affected by smooth changes in level, whereas Model III provides further flexibility, allowing for changes in slope[8].

The LM statistic to test the null of stationarity has the expression:

$$\hat{S}_T=\hat{\sigma}^{-2}T^{-2}\sum_{t=1}^{T}E_t^2 \quad (9)$$

with $E_t=\sum_{i=1}^{t}e_i$ being the forward partial sum of the residuals of nonlinear least squares (NLS) fitting, and $\hat{\sigma}^2$ being a suitable estimator for the long-run variance of $\{\varepsilon_t\}$.

### 2.3. Estimation of change locations

Knowledge about stationarity of the time series is useful for accurate estimation of the change dates. More precisely, Kejriwal and Lopez (2013) concluded that, when a unit root is present, more accurate estimates for the break dates can be obtained by estimating a specification in first differences, while in the $I(0)$ case better results in terms of mean squared errors are obtained when estimating a level model. The estimated change dates allow us to determine the various regimes for $\beta$-convergence analysis below.

### 2.4. $\beta$-convergence analysis

The general idea of convergence from a time series perspective (*e.g.* Carlino and Mills, 1993; Quah, 1993; Bernard and Durlauf, 1995; Evans and Karras, 1996; Li and Papell, 1999) has resulted in several, more specific, concepts of convergence. In this paper we will focus on so-called *stochastic convergence* (or *catching-up*), which refers to the case when the logarithm of the relative series is stationary around a deterministic trend. Given that the presence of a time trend allows for permanent differences we conclude, following

[8] Models I and III are estimated by nonlinear least squares (Levenberg-Marquardt algorithm with preliminary grid search is applied, as in Presno *et al.*, 2014).

Carlino and Mills (1993) and Tomljanovich and Vogelsang (2002), that stochastic convergence is attained when both stationarity around a trend and $\beta$-convergence hold. This way, shocks in relative per capita $CO_2$ emissions should be temporary but -in addition to this- if a country has initial emissions above the mean, the subsequent rate of growth of its emissions should be negative for convergence to occur. Conversely, if the country initially is below its compensating differential, its subsequent rate of growth should be positive. In the linear framework this implies that, if per capita emissions are converging, a regression of the logarithm of relative per capita emissions on the intercept and a linear trend should have opposite signs in their estimated coefficients. In this paper we extend the above methodology to the quadratic case, with a separate treatment of break and smooth transition models.

In the break case we estimate the following equation for each country:

$$y_t = \beta\left(\frac{t}{T}\right)^2 + \sum_{k=1}^{n+1}\delta_k BDU_{k,t} + \sum_{k=1}^{n+1}\eta_k BDT_{k,t} + u_t \qquad (10)$$

where $BDU_{k,t} = 1$ and $BDT_{k,t} = (t - T_{b,k-1})/T$ for $T_{b,k-1} < t \leq T_{b,k}$, and $BDU_{k,t} = BDT_{k,t} = 0$ otherwise, with $T_{b,k}$ denoting the $k$-th change point estimated in Subsection 2.3 and $k$ = 1,..,$n$ (where $T_{b,0} = 0$, $T_{b,n+1} = T$, and $n$ is the number of changes detected in the first stage[9]).

In the smooth transition study we have the following model:

$$y_t = \beta\left(\frac{t}{T}\right)^2 + \sum_{k=1}^{n+1}\delta_k SDU_{k,t} + \sum_{k=1}^{n+1}\eta_k SDT_{k,t} + u_t \qquad (11$$

where $SDU_{1,t} = 1 - sigm_{1,t}$, $SDU_{k,t} = sigm_{k-1,t} - sigm_{k,t}$, $2 \leq k \leq n$; $SDU_{n+1,t} = sigm_{n,t}$, with $sigm_{k,t} = [1 + \exp\{-\gamma_k (t/T - \lambda_k)\}]^{-1}$, for $1 \leq k \leq n$, $1 \leq t \leq T$ ($\lambda_k = T_{b,k}/T$ and $\gamma_k$ are, respectively, the relative position of the timing of the transition midpoint and the speed of transition, both estimated previously). Finally, $SDT_{1,t} = (t/T)SDU_{1,t}$ and $SDT_{k,t} = (t/T - \lambda_k)SDU_{k,t}$, $2 \leq k \leq n+1$.

In the linear framework, Tomljanovich and Vogelsang (2002) study $\beta$-convergence by analysing the signs and significance of coefficients $\delta_k$ and $\eta_k$ in each regime, interpreting that convergence occurs when the estimates for both parameters have opposite signs. Convergence assessment is more delicate in the quadratic model, as the presence of the quadratic term –as well as the smooth/instant transition component- implies that a naïve analysis of the signs of $\hat{\delta}_k$ and $\hat{\eta}_k$ is not enough for convergence, as the rate of growth of emissions is a relatively complex nonlinear function of time depending on all the parameters of the model. The problem is readily solved –in the smooth transition case- by a suitable generalization of the rule proposed by Tomljanovich and Vogelsang (2002). The idea is straightforward: it is the sign of the rate of growth, rather than just the sign of $\hat{\eta}_k$, what determines convergence. As that rate is not constant in the quadratic case –it evolves in time, as it depends on both $t^2$ and the logistic components of the model- some kind of summary indicator is necessary. A simple idea is to calculate a suitable average (*e.g.*, the mean or median) of the time derivative of the deterministic part of model (11). Here we opted for median derivatives[10] as they are less sensitive to extreme values. Then the signs of the estimated function (11) at the starting point in each regime and that of the median derivative of the estimated function in the same regime are compared. If both have opposite signs, then convergence would occur.

[9] Carrion-i-Silvestre and German-Soto (2009) adopt this procedure -for the linear trend case- in their study of stochastic convergence among Mexican regions.

The same idea is applied in the quadratic model with breaks –strictly, one-sided derivatives are used in this specific setting-, and the signs of $\hat{\delta}_k$ and the median right derivative of the estimated function in each regime are compared. In the linear case this procedure coincides with the rule proposed by Tomljanovich and Vogelsang (2002).

Summing up, we consider that $\beta$-convergence is attained when the sign of $\hat{\delta}_k$ or that of the value of the estimated function at the beginning of the regime (for the break or smooth transition models, respectively) and that of the median derivative of the estimated function in that regime are opposite. Following the nomenclature proposed by Tomljanovich and Vogelsang (2002), in the results of Section 3 below convergence will be denoted as *C* or *c*, respectively, depending on whether the coefficients of both parameters are significant at 10% level or it is only one of them that is significant. Following the same criterion, divergence is denoted by *D* or *d*, respectively, and is meant to occur when both signs coincide. We will also denote by *E* the case when estimates are very small in magnitude and statistically insignificant, suggesting $\beta$-convergence (equilibrium growth).

[10] Most conclusions in our analysis do not change depending on the criterion (median/mean derivative) considered.

## 3. EMPIRICAL RESULTS

This section includes an empirical analysis of stochastic convergence. In order to compare results[11] we consider the group of countries studied by Westerlund and Basher (2008): 28 OECD countries, classified in developed and developing states. The list of the former includes Australia, Austria, Belgium, Canada, Denmark, Finland, France, Germany, Italy, Japan, the Netherlands, Spain, Sweden, Switzerland, the United Kingdom, and the United States. The developing countries are Argentina, Brazil, Chile, China, Greece, India, Indonesia, Mexico, New Zealand, Peru, Portugal, and Taiwan.

We carried out the analysis for the whole sample of countries (joint analysis) and also separately for developing and developed countries (group analysis). That distinction is relevant as, among other reasons, developed countries have experienced a change of economic focus -mainly from the industry to the services sector- and this process is expected to lead to a reduction in emissions. However, developing countries have growing industrial sectors whereas the weight of agriculture tends to reduce. Also, developed countries have outsourced their most highly pollutant industries, now installed in developing states. In addition, some studies indicate that if convergence were achieved in developed countries, this would encourage developing countries to accept a cap on their own emissions[12].

In order to obtain a homogeneous sample we studied the 1901-2009 period, although for some countries data are available since year 1870. National data on $CO_2$ emissions (in metric tonnes) come from the Carbon Dioxide Information Analysis Centre[13] (CDIAC), and reflect anthropogenic emissions from fossil fuel consumption, cement manufacturing and gas flaring. The population data were extracted from Maddison (2010).

[11] Westerlund and Basher (2008) study convergence using Evans (1998) definition. Here we shall follow the definition proposed by Carlino and Mills (1993).

[12] For instance, the US Congress refused to allow ratification of the Kyoto Protocol until major developing countries like China and India commit themselves to reduce their own GHG emissions in coming years.

[13] Source: Boden *et al.* (2012).

Following Carlino and Mills (1993) analysis of stochastic convergence in per capita income, we compute the logarithm of the ratio of per capita emissions in each country to average per capita emissions of the set of countries. Three cases are considered, depending on whether the average of the whole sample or only that of the developed or developing countries is taken as the reference basis for the analysis. Therefore, the variable of interest is relative per capita emissions. Nonstationarity in the logarithm of that series would support divergence since the effects of a shock are permanent and there is no tendency for per capita emissions to converge to the average. However, stationarity would imply a mean reverting behaviour. As Camarero *et al.* (2011) remark, given that Carlino and Mills (1993) definition assumes nonstationarity[14] of the individual series, it is crucial to carry out a stationarity analysis of the logarithms of $CO_2$ per capita emissions previously to any assessment on convergence.

Following the methodology described in Section 2, our study begins with this type of analysis. In the first stage we applied statistical tests to ascertain if breaks are present. In order to distinguish between changes in level and slope, we followed the strategy applied by Presno *et al.* (2014), based on the sequential procedure of Kejriwal and Lopez (2013). It begins by testing for one structural break under Model III, using the procedure proposed by Perron and Yabu (2009). If the null is rejected, this may be due to a change in level and/or slope, so the unrestricted test is applied in the second stage. A rejection by this test can be attributed to a change in the growth rate. Once evidence in favour of a break is detected, we test for one versus two breaks by the Kejriwal and Perron (2010) procedure, extended to the quadratic case. Below we shall report the results of the one versus two breaks test -regardless of the conclusions of the single break test- because of potentially low power in the presence of multiple breaks, mainly when consecutive changes have opposite signs. Finally, conditionally on a stable slope in the first step, we study the number of level breaks by using Harvey *et al.* (2010) test.

Taking into account the relatively short length of the series, we allowed for a maximum of two changes in our analysis. Kejriwal and Perron (2010) recommend deciding the maximum number of breaks with regard to sample size in order to avoid a small number of observations in each subsample, with consequent potential problems in the performance of the tests[15]. Also, if the objective is to study the presence of a unit root in the series, allowing for a large number of breaks may be problematic as a unit root process can be viewed as a limiting case of stationary process with multiple breaks (*e.g.*, Kejriwal and Lopez, 2013).

[14] Applying Carlino and Mills methodology involves defining the log ratio of two series ($CO_2$ per capita and average $CO_2$ per capita emissions). If they both are stationary, their linear combinations would also be stationarity, and Carlino and Mills's measure would be meaningless.

Finally, we conduct the analysis considering two possibilities for the change, namely, an abrupt change (break model) and a gradual change (smooth transition model) between regimes.

**Analysis of per capita $CO_2$ emissions**

Table 1 below reports the results[16] of stationarity analysis for the logarithm of per capita $CO_2$ emissions, prior to the assessment of convergence. At 10% significance level, the null of stationarity is rejected for most series in at least one model (smooth/break) class. Exceptions are two series -Canada and Chile- for which no change was detected and Spain. Meanwhile, conclusions for Argentina, China, Finland, Greece, Indonesia, the Netherlands and the United Kingdom series depend on the model considered and, with the exception of Greece, specification of a smooth change leads to rejecting the null of stationarity. In these problematic cases, use of model selection criteria to decide between break and smooth models led us to select smooth transition specifications. Therefore, we conclude nonstationarity of the individual series. Thus, the effect of a shock in per capita emissions would be permanent, and the series would not revert to its mean value. This finding has practical consequences on emissions forecasting, an issue that worries both scientists and policy makers.

[PLS. INSERT TABLE 1 ABOUT HERE]

It is also noteworthy that for many countries both kinds of models lead to similar change points. For most of the developed country series -mainly the European ones-, the procedure detects changes in periods about the two Great Wars.

[15] This strategy is frequently used in empirical analysis. For instance, Kejriwal and Lopez (2013) analyse per capita output on samples of 137 observations, justifying the inclusion of a maximum of 2 breaks.

[16] For brevity Table 1 reports stationarity analysis for a specific value of the user-supplier constant ($k$ = 0.9) required for the data-driven device to compute the bandwidth. Notwithstanding, similar results and identical conclusions were obtained for $k$ = 0.5.

A change around World War I is detected in Argentina, Austria, Denmark, Finland, Greece, Mexico[17], the Netherlands, Portugal and Sweden[18].
World War II could affect Germany, Italy (the Allied Invasion of Italy took place in 1943), Japan (1945 is the year of the surrender of the Empire of Japan), the Netherlands, Sweden[19], Switzerland, and Greece[20]. Meanwhile, France and Belgium show similar features[21], and the 1946-1947 biennium matches the rebuilding of Europe after World War II. In Austria the break model detects a change in year 1938, when the German troops entered Vienna and Austria was annexed to the Third Reich.
Certainly, the above events had a significant weight on per capita $CO_2$ emissions in the OECD countries, since the series of averages -for developed, developing, and the whole group of countries- show changes around 1914 and 1944.
Another event that seems to have affected the per capita $CO_2$ series was the Great Depression. This is the case of Australia, Mexico[22], New Zealand, Peru and the United States.
Also, the models detect changes that might be related to specific events affecting each country, as the Civil War (1936) in Spain, the war between the Soviet Union and Finland (1939), the General Strike of 1926 in the United Kingdom, the Recession of 1949 in the United States, and the Brazilian re-democratization period (widely known as the Second Republic) that began in 1946. In the case of China, the models detect a change in 1907 that might relate to the Chinese Famine -the second worst famine in recorded history, with an estimated death toll of around 25 million people-; also, 1955 was the central year of the First Five-Year Plan for Development of the National Economy of the People´s

---

[17] Two relevant events influencing 20$^{th}$ century Latin America were the Mexican Revolution and World War I. In spite of neutrality of most Latin American nations, WWI affected them as a consequence of the interruption of European demand for their products; also, some local industries began to produce replacements for European products.
[18] Despite its neutrality in the First World War, 1917 was a hard year for the Swedes due to maintenance problems and the effects of the Russian Revolution.
[19] Neutrality did not preclude Sweden from being affected by the war between the Soviet Union and Finland in 1939 and the German invasion of Norway in 1940. The estimated midpoint change date, in year 1945, could be related to the end of World War II. Switzerland is another country that, despite its neutrality, was affected by the War.
[20] Notwithstanding Greek neutrality, Italian troops crossed the border on the 28th of October 1940, beginning the Greco-Italian War.
[21] Both countries show a change around 1923, a period that matches the French and Belgian Occupation of the Ruhr.
[22] Following the 1929 economic crisis several countries in Latin America entered a new historical stage as they could not export raw materials and metals or import manufactured products from Europe and the United States. So, they turned to industrialization, and the number of industries in countries like Mexico extended considerably.

Republic of China (1953-57), whose declared objective was striving for economic growth and emphasizing development in heavy industry and technology rather than agriculture.

The above changes in country emissions would directly relate to changes in output. Concretely, Li and Papell (1999) analyse convergence in per capita output for 16 OECD countries, finding that World War II affected Austria, Belgium, Canada, Denmark, France, Germany, Italy, Japan, the Netherlands, Switzerland, and the United States; breaks also occurred in the 1920s in Australia (related with the industrial expansion after World War I), Finland (explained by its independence from the Soviet Union and the subsequent civil war), Sweden and the United Kingdom (after the chain of strikes culminated with the general strike of 1926). Ben-David and Papell (1995) also find that the Great Depression was the cause of breaks in Canada and the United States. Conclusions along the same lines are also obtained by Kejriwal and Lopez (2013), applying a procedure similar to ours.

**Analysis of convergence in relative per capita emissions**

Once concluded that most of the series are nonstationary, we analyse convergence in relative per capita emissions, defined as the logarithm of the ratio of per capita emissions in each country to average per capita emissions of the set of countries.

Table 2 reports the number of changes estimated according to the procedure described above, for the whole set of countries (joint analysis) and separately for developed and developing states (group analysis). Column "Model- # changes" indicates the model and the number of changes selected according to the sequential procedure (Model 0: no changes; Model I: change in level; Model III: change in both level and growth rate)[23].

[PLS. INSERT TABLE 2 ABOUT HERE]

Summing up, Model III with two changes is selected in most cases, and zero structural changes are only detected in Australia and Denmark (joint analysis) and Australia (developed countries set). Also, conclusions about both model and number of changes do not vary significantly when joint and group analysis are compared, particularly among developed countries. Only some differences are observed for Canada, Denmark, Argentina, India, New Zealand and Taiwan.

[23] In the problematic cases (namely, Austria, the Netherlands, Brazil, Chile, and India) when the null hypothesis of stability is rejected for the "general" specification but not for the "unrestricted" one, and Harvey *et al*. (2010) test detects no changes, we opted for Model III in order to avoid size distortions.

Tables 3 and 4 summarize the results of stationarity testing on relative per capita $CO_2$ emissions, for joint and group analysis, respectively. We report results for both the smooth transition and the break case, with[24] $k$=0.5 and $k$=0.9.

[PLS. INSERT TABLES 3 AND 4 ABOUT HERE]

In joint analysis, the null of stationarity is not rejected (under both smooth and break models) at 5% significance level for Belgium, the Netherlands, Argentina, Brazil, and Mexico (as well as Australia and Denmark under the no change model). Also, if the smooth model is considered, Canada, France, the United States and Greece series would be stationarity. Under the break specification, the list of stationary series would be expanded with Germany, Japan, Spain, the United Kingdom, Chile, China, Indonesia, and New Zealand (see Figure 1).

[PLS. INSERT FIGURE 1 ABOUT HERE]

As for group analysis, stationarity is not rejected neither for smooth nor for break models (at 5% significance level) for Belgium, Denmark, France, Germany, the Netherlands, Spain, Argentina, Indonesia, and Mexico. The list would be enlarged with Canada, Chile, Greece and India (only under the smooth transition model), as well as Japan, Brazil, China and Peru (only under the break specification) and Australia (under no change model) (See Figure 1).

It is clear that for some series (Austria, Finland, Italy, Sweden, Switzerland, Portugal, and Taiwan), the null of stationarity is rejected under all the (smooth, break) models and approaches (joint, groupwise) considered. Thus, no evidence of convergence is detected in those countries.

As conclusions may vary in some cases depending on the specific kind of change (smooth/abrupt), Tables 3 and 4 also report the results of model selection criteria (Schwarz´s information criterion –SIC-, Akaike´s information criterion –AIC-, and adjusted $R$-squared). These criteria only select abrupt change models for Germany, China and Greece (Tables 3 and 4), Mexico (Table 3) and Spain (Table 4). This would be consistent with the general existence of gradual -rather than abrupt- changes in the process of potential convergence in relative per capita emissions of $CO_2$.

Under the models selected by the above criteria, and at 5% significance level, the null of stationarity is not rejected in Australia, Belgium, Canada, Denmark, France, Germany, the Netherlands, the United States, Argentina, Brazil, China, and Mexico (Table 3). By country groups (Table 4), the stationary series would be Australia, Belgium, Canada, Denmark, France, Germany, the Netherlands, Spain, Argentina, Chile, China, India, Indonesia, and Mexico.

[24] Similar results were obtained for $k$=0.8.

Results of stationarity testing allow change points to be estimated for the above series. In an integrated process the change points shall be estimated on the series in first differences, whereas estimation in case of stationarity should be carried out on the level of the series instead. Tables 5 and 6 (for joint and group analysis, respectively) include the estimated change dates for both smooth and break models, revealing differences in some cases. These estimated change dates also allow us to identify the various regimes for $\beta$-convergence analysis. For series being stationary around a trend, we analyse the $\beta$-convergence condition by estimating equations (10) and (11) for both the break and the smooth transition models. Tables 5 and 6 report the estimated coefficients for each regime and associated significance tests, using Newey and West (1994) robust covariance matrix estimates. It is worth highlighting that the quadratic trend is statistically significant for most series, confirming their nonlinear nature.

[PLS. INSERT TABLES 5 AND 6 ABOUT HERE]

**Results of *β*-convergence analysis**

For those particular series identified as stationary under both smooth and break models, the conclusions of $\beta$-convergence analysis for the last regime are remarkably robust, not being affected by the specific model selected (excepting the cases of France and the Netherlands in group analysis). Now we comment the results separately for joint and group analysis. (Figure 2 below also includes a sample of some representative cases).

[PLS. INSERT FIGURE 2 ABOUT HERE]

Attending to joint analysis for the OECD countries, and taking the last regime as a reference, we may note the following findings:

a) All the countries classified as stationary tend to convergence to the global average, that is, those members exceeding the average have negative median derivatives whereas those under the average display positive ones. Evidence of so-called *divergence from above* is only found for a single series -Australia- where no changes are detected. It is worth noting that coal is the main source of electricity generation in Australia (which is one of the main exporters of coal in the world), although we must also take into account the existence of a unique regime (*i.e.*, a very long time span) in that specific series.

b) Different behavior patterns are observed between developed and developing countries. At the beginning of the last regime, the majority of developed countries emit $CO_2$ above average, while emissions of developing nations kept below the mean. Otherwise said, developed countries -with the exceptions of Denmark, Japan and Spain (the latter two under break models)- converged from above, whereas convergence from below is observed in developing states.

Japan and Spain started from very low levels of industrialization at the beginning of their last regime, which would coincide with the end of World War II and the Spanish Civil War, respectively. Afterwards, they experienced a great development (*the Japanese post-war economic miracle*, and the Spanish industrial and tourist boom in the 1960s), that resulted in positive median derivatives in their convergence paths, due to strong GDP per capita effect. For the Spanish case, this would be in accordance with conclusions of Fernández González *et al*. (2014a). In the Japanese case, Schymura and Voigt (2014) observe declines in emissions since 2007 (with reference in year 1995), caused by the economic crisis; in addition, Japan has made a major effort in recent years in order to change its sectoral composition towards less polluting sectors.

Denmark and Australia are the only countries where no changes are detected in their relative emissions. Given the long period considered a comparison with the above cases is not direct.

c) Interestingly, some OECD members show negative median derivatives for the last regime. These include Belgium, Canada, France, the Netherlands, Germany, the United Kingdom, and the United States. Also, with the exceptions of Germany and the United Kingdom under the break model, the last regime begins around 1946/1947, coinciding with the rebuilding of Europe after World War II.

Belgium and France exhibit very similar patterns; in fact, both series appear almost parallel, with the Belgian series being above the French one. After the War, both these countries, along with the Netherlands, saw an era of great economic growth, with the rebuilding of national social institutions and industry and the process of European integration[25] that changed the continent permanently. From the 1950s on we observe negative derivatives in their estimated functions, which would match their efforts to control emissions. Fernández González *et al.* (2014a) observe reductions in Belgian emissions in recent years, as the country has promoted renewable energy and energy efficiency measures. Recent studies in the field of index decomposition analysis also highlight the reductions in French total emissions, as a consequence of efforts in fuel mix (power generation increases based on less carbon intensive energy sources; in this regard the country's commitment to nuclear energy is also worth remembering), structural change (shifts in sectoral composition towards less polluting sectors), and intensity effects (declines in sectoral carbon emissions based on energy intensity, or rise in energy efficiency), which have tended to offset the expansive effect of economic growth (activity effect) on $CO_2$ emissions.

[25] The European Coal and Steel Community (1951) and the European Atomic Energy Community (1957) are created in that time, with France having a relevant weight on it.

The French and Dutch series show similar behavior and levels until the 1960s, although the former seems to have had a greater control on emissions since then. In any event, we observe negative derivatives in the estimated Dutch function from the 1970s on. As in the French case, Fernández González *et al*. (2014b) and Schymura and Voigt (2014) show reductions in Dutch total emissions as a consequence of fuel mix, structural change and intensity effects.

The Canadian and US series also exhibit similar patterns, with the level of the latter exceeding that of the former although tending to get closer during the most recent regime, beginning in year 1947. Since then, the time derivatives of the estimated functions have been negative in both cases, with the American series generally exhibiting higher absolute values. The Canadian economy boomed during the war as its industries manufactured military material for other countries. Thus, Canada finished the war with a large army and a strong economy. Those circumstances would explain the change detected around 1947. In recent years, Schymura and Voigt (2014) highlight that the combination of a slight decrease in the use of carbon intensive energy sources, as well as a significant structural change towards less polluting sectors and improvements in energy efficiency, have been dampening the effect of economic growth. Meanwhile, The United States is doing a relevant effort in favor of “cleaner” sectors that are gaining weight in its economy.

For the United Kingdom the change (under the break model) is located in 1976, coinciding with the discovery of oil in the North Sea and recovery from the crisis. In the German case the most recent regime begins around 1952-1953, coinciding with crucial events in recent German history such as the signing of the Bonn–Paris conventions -putting an end to the Allied occupation of West Germany- and the Treaty of Paris -that established the European Coal and Steel Community- with the objective of unifying countries after World War II.

Finally, some remarkable cases are New Zealand, China and Chile. New Zealand shows values of both its time series and the derivative of its estimated model that are near zero for the last regime. This is no surprise as that country achieved the first place in the Environmental Performance Index ranking in year 2006 (Yale University).

The establishment of the Popular Republic of China in 1949 and the aforementioned First Five-Year Plan for Development of the National Economy (1953-57) could be the germ of the take-off and strong economic growth achieved by the country, especially since the late 1980s. All studies highlight the tremendous expansive effect of the economic growth impact (activity effect) as the cause of the huge increase in Chinese emissions of $CO_2$, which accounted for around 25% of the world's emissions in year 2010. Indeed, a positive derivative is observed in the estimated function throughout the last regime.

Chile shows convergence from above and a recent change point in year 1988. That would coincide in time with the end of the dictatorship, the beginning of the democratic transition and opening to the outside world[26]. A main contributor to $CO_2$ emissions in Chile is the mining sector, which is one of the pillars of the economy. In fact, that country produces over 33.3% of the global copper output, and -as a consequence of the increase in production and ageing mines- fuel consumption by the copper mining industry increased by 66% between 2001 and 2012. That consumption has a direct relationship to GHG emissions.

Regarding convergence analysis by groups (developed and developing countries), we find some interesting results. Some changes compared to the joint analysis are observed. For instance, the New Zealand, British and North American series are now classified as nonstationary. Conversely, the Peruvian (under abrupt change specification) and Indian series now add to the group of stationary series.

As a result of higher average emissions in developed countries, the series of relative per capita emissions for those countries show lower levels than in joint analysis. The contrary occurs for the developing countries:

a) First, regarding developed countries, GDP per capita (OECD data) does not seem to be the sole determinant of country emissions being above or below their group average. Other factors like innovation, search for more efficient technologies and fossil fuel substitution (*e.g.*, the cases of France and Germany) would also be essential. Also, unlike developing countries, the change dates for developed states generally coincide with those from joint analysis.

[26] Derivatives are positive since late 1980s, coinciding with the opening of the Chilean economy.

Developed countries show bigger dispersion in their convergence patterns than developing ones. Specifically, we find divergence in three of the nine stationary series under some (break/smooth) specification. France and Germany diverge from below and Canada from above in the last regime. The other countries converge: Belgium and Denmark from above, while Australia, Japan, the Netherlands and Spain converge from below.

The German and French cases are particularly interesting as, while both of them are below their group average and diverge, they show negative median derivatives in their emissions. In the French case the derivative keeps negative throughout all the last regime, whereas it became positive in Germany by the beginning of the 1990s (this could be related to the German reunification). As a difference between both countries, while values are increasingly negative in the French series, they oscillate around zero in the German series. Growing environmental awareness, development of clean renewable energy, use of more efficient technologies, and environmental taxation, made that possible (Fernández González *et al.*, 2014b).

Conversely, if the median[27] derivative criterion is applied, Canada would diverge from above. We observe a positive derivative in its estimated function from the end of the 1970s. It is also remarkable that Canada is one of the world's highest per capita consumers of energy -pushed up by its cheap energy- as well as one of the few developed nations being a net energy exporter.

The change in the Danish series occurs around the early 1960s. The OECD's Economic Survey of Denmark (1962) highlights the accelerated structural change that took place in those years. Since then, Denmark has a negative derivative in its estimated function. A similar behavior is observed in the Belgian series from the 1950s on.

Convergence for Australia, Japan, the Netherlands and Spain is from below. Although beginning from negative levels, they all attained positive median derivatives; however, while in the Spanish case the derivative of the estimated function was positive throughout all the last regime, in the Japanese[28] and Dutch cases we observe negative derivatives from late 1990s and the 1980s, respectively.

---

[27] It must be noted that the mean derivative criterion would lead to convergence from above in this case.

[28] In the Japanese case this might relate to slow growth during so-called *Lost Decade* after 1990.

b) Secondly, in relation to the developing country group analysis, those nations show low dispersion in their convergence paths. Also, their change points do not seem to be linked to wars, but specific events related to their economies and political situation appear to be more clear determinants.

Most countries in the group converge: from below in the cases of China, Greece, Indonesia, Mexico, Chile, and India, and from above in Argentina. However, we find two cases -namely, Brazil and Peru- where divergence from below is detected for the break model.

In the Greek case, if the smooth change model is considered, the last regime begins in 1948, coinciding with the last years of the Greek Civil War. Fernández González *et al.* (2014a) highlight the increase[29] in emissions in recent times, due to the effect of GDP per capita and high carbon contents of several energy carriers (carbonization effect), that could not be neutralized by efforts in innovation, search for more efficient technologies, and fossil fuel substitution.

The Indian series shows a quite recent change in year 1994 that might relate to the economic liberalization beginning in 1991, which brought about a large increase in GDP. Thus, the huge rise in emissions in India seems to be due to its strong activity effect. Meanwhile, Indonesia witnessed economic and political changes in the 1960s that finally led to the economic boom of that *Asian tiger*. For the 1995-2009 period, Schymura and Voigt (2014) remark the large increase in emissions, mainly due to the activity effect and the substantial structural change towards more carbon intensive sectors.

The Mexican series also converges; however, the initial level of emissions in that country was very low since the change date is fairly old; since then, Mexico has experienced strong growth that seems to be the responsible for the increase in emissions (Schymura and Voigt, 2014).

We find three cases where median derivatives are negative in the last regime: Argentina, Brazil, and Peru. This could be motivated by strong growth in emissions in other countries included in the group of developing states. However, while in the case of Argentina we find convergence from below, Brazil and Peru would diverge from below. A possible explanation for that difference is that the turning point in Argentina was quite early, in year 1915, and Argentinian emissions at that time exceeded the average of other developing countries (until the 1930s, the country's per capita income was similar to that of France, Germany and Canada). However, the derivative of the estimated function becomes negative from the end of the 1950s. Meanwhile, Brazil and Peru have always been below the average, and their derivatives are negative –yet quite close to zero- from the 1960/1970s. In most recent years, Schymura and Voigt (2014) highlight the increase in Brazilian emissions as a consequence of substantial economic growth, combined with use of less energy efficient technologies.

[29] A slowdown is observed in that country -with negative derivatives in its estimated function- since the end of XXth century.

## 4. CONCLUSIONS

Most studies on convergence in $CO_2$ emissions have been carried out within a linear framework. However, emissions appear to be related to income, and some researches have emphasized the nonlinear nature of economic activity as well as the convenience of considering changes that occur gradually -rather than instantaneously- in economic systems. In this paper we have addressed convergence analysis for 28 OECD countries from a nonlinear standpoint, considering quadratic trends and allowing for smooth transition specifications in addition to commonly used break models.

The analysis has begun with a study of individual series of per capita $CO_2$ emissions, concluding that most of them are nonstationary. This would imply that shocks have permanent effects, which has practical consequences on emission forecasting and should be taken into account by scientists and policy makers. Also, the most relevant change dates in European countries during XXth century locate around World Wars I and II and the rebuilding after them. Meanwhile, Australia, Mexico, New Zealand, Peru and the United States were strongly affected by the Great Depression, whereas the changes in other countries are related to more idiosyncratic aspects of their political and economic development (*e.g.*, the Civil War in Spain and the General Strike of 1926 in the United Kingdom). Most changes in the per capita $CO_2$ series clearly coincide with changes detected in per capita output (*e.g.*, Li and Papell, 1999, and Kejriwal and Lopez, 2013), due to the strong relationship between both magnitudes.

Stationarity analysis for per capita and relative per capita $CO_2$ emissions was carried out by considering both instant and smooth changes, with non-coincident conclusions obtained in some cases. Model selection criteria allowed us to conclude that smooth transition models would be more suitable than break-based specifications in most cases. Hence, in the bulk of the series considered in this study, changes would be gradual rather than instantaneous. In addition, although our analysis indicates that almost all the above per capita $CO_2$ emission series would be nonstationary, stationarity is also detected for many relative per capita emission series (this being a necessary but not sufficient condition for stochastic convergence). More specifically, stationarity in the relative emissions series is only rejected -regardless of the specific (break, smooth transition) model considered- for Austria, Finland, Italy, Sweden, Switzerland, Portugal, and Taiwan.

Convergence analysis was addressed from a twofold perspective: first joint study of the 28 OECD countries, and then group analysis -distinguishing between developed and developing countries, following the classification of Westerlund and Basher (2008)- with stationarity testing and $\beta$-convergence analysis adapted to the quadratic framework.

Both joint and the group analyses find -for the last regime- negative median derivatives of the estimated relative emissions function in many developed countries, whereas the same figures are positive for almost all developing countries. In addition to factors like search for innovation, concern with the environment, fossil fuel substitution, and the expansive effect of the increase in GDP, an issue to consider is that developed countries have experienced a shift from the industry to the services sector, also outsourcing industries that stand out for their higher emissions. Meanwhile, developing countries have increased their industrial sectors to the detriment of the agriculture.

Specifically for the joint analysis we conclude that, with the exception of Australia, all those series -totaling nineteen cases- classified as stationary under some of the (break/smooth transition) specifications considered would converge in their most recent regimes. However, in group analysis we find a slightly smaller number of eighteen -nine in each group- stationary series, in addition to higher dispersion in terms of $\beta$-convergence. More precisely, in the developed group and for the last regime, three among the nine series diverge: France and Germany from below (beginning from lower levels than the other members of the group and exhibiting negative "growth rates"), whereas Canada –under the specific criterion considered in our study- diverges from above. Also, we must highlight the cases of Belgium and Denmark, which converge from above. It appears that GDP per capita is not the sole determinant of convergence in that group, with other factors -like search for more efficient technologies, fossil fuel substitution, innovation, and perhaps outsources in industries- also being key issues.

Smaller dispersion in their convergence processes is detected in developing countries. Most of them (China, Chile, Greece, India, Indonesia, Mexico) would converge from below, and only Brazil and Peru diverge from below. The former six countries have experienced a strong growth that seems responsible for a large part of the increase in their emissions. Indeed, they have also been receptors of high-emission, outsourced industries. In short, our analysis suggests that countries are not following completely independent paths in pollution control, but are instead moving towards a common standard of environmental performance. In this regard, an interesting research avenue would suggest to extend our study in order to encompass stronger concepts of convergence, such as *deterministic convergence* (implying that relative emissions series are stationary around a non-zero level) and *absolute*, *unconditional* or *long-run convergence* (requiring zero mean stationarity).

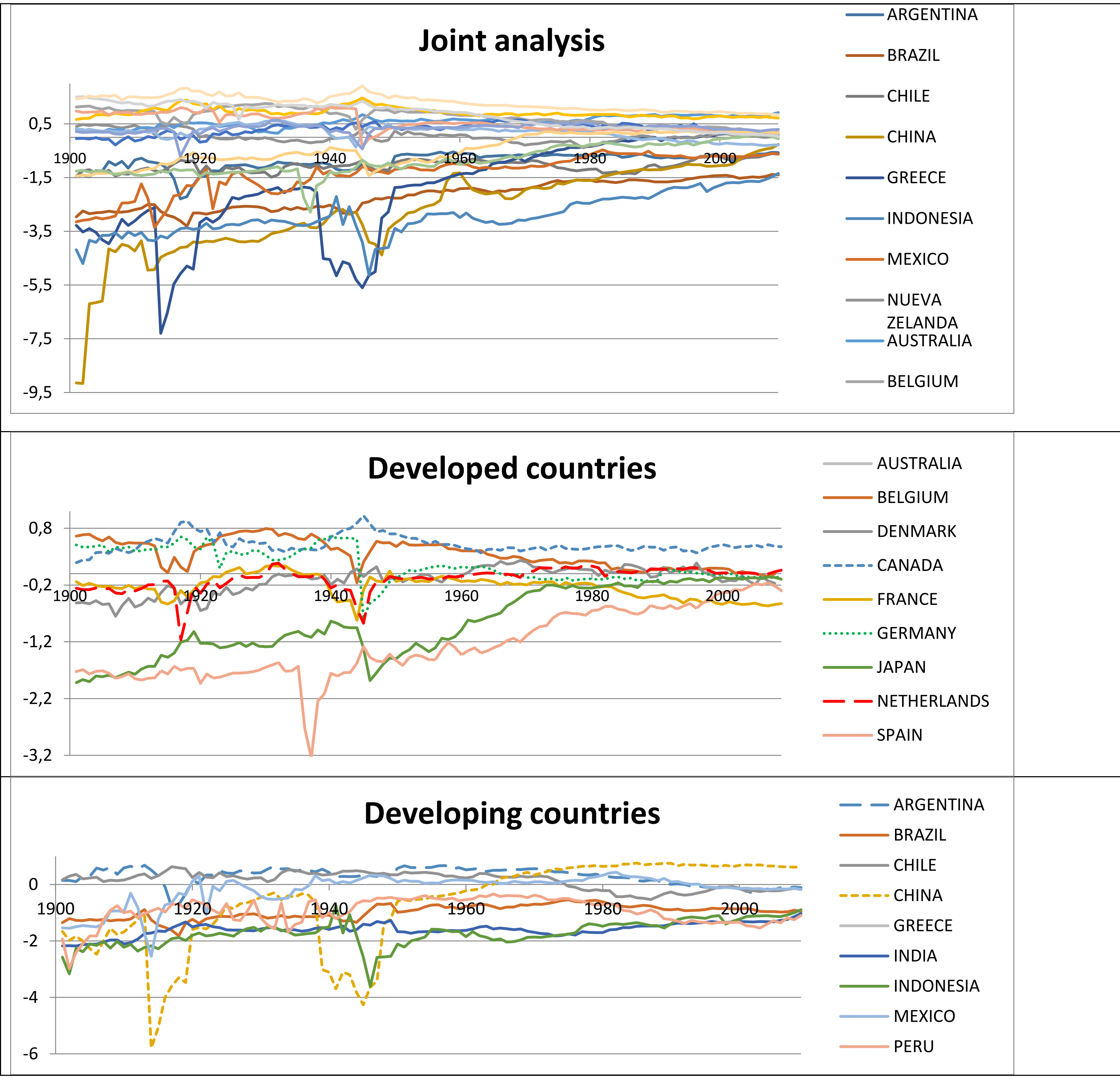

**Figure 1**. Joint and group analysis. Time paths of relative per capita $CO_2$ emissions for those series classified as stationary under at least one (break/smooth transition) model specification.

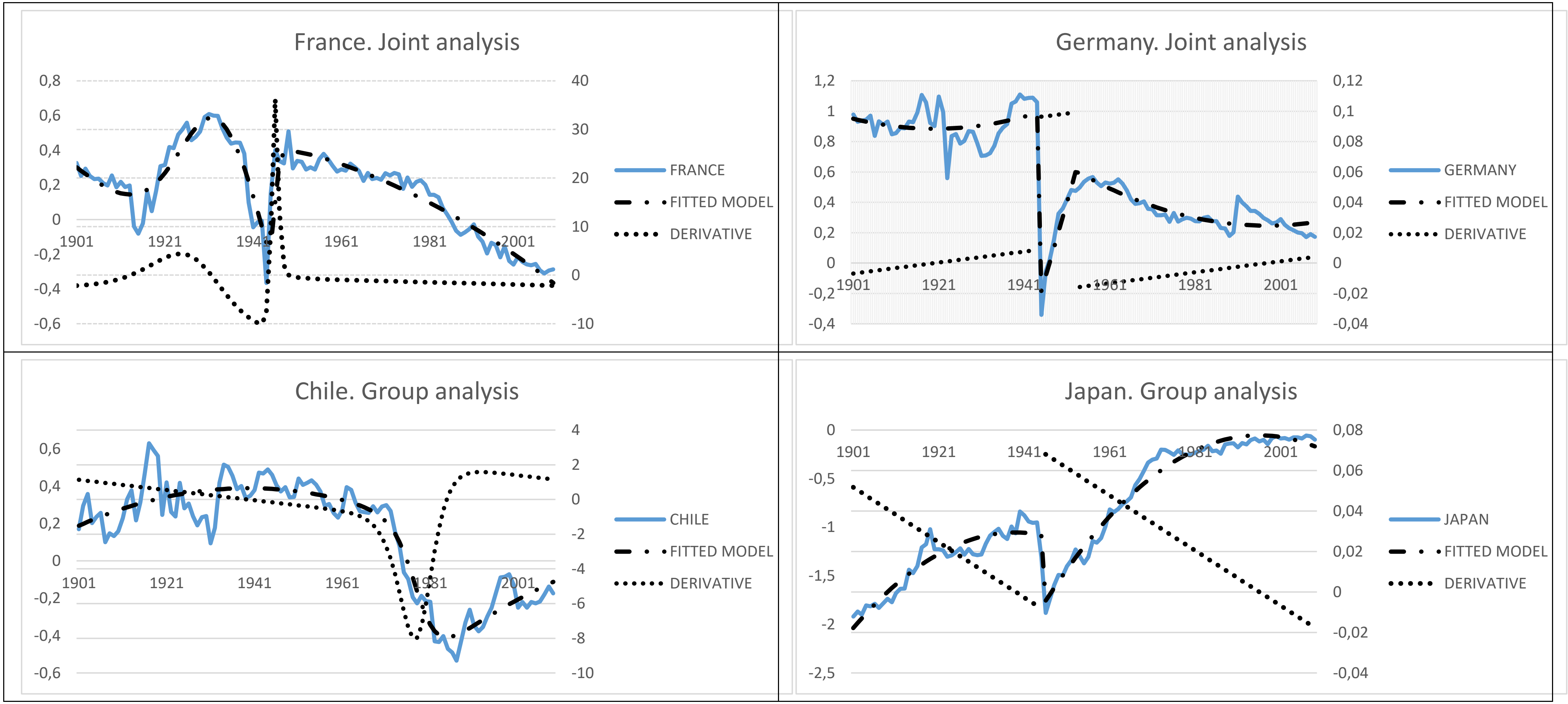

**Figure 2**.Time series/fitted models (left axis) and derivatives (right axis)

**Table 1. Results of stationarity testing on logarithms of $CO_2$ per capita emissions. Smooth transition and break models**

| DEVELOPED COUNTRIES | | | DEVELOPING COUNTRIES | | |
|---|---|---|---|---|---|
| | Smooth model | Break model | | Smooth model | Break model |

| Country | Model | Estimated parameters and midpoint change date | Statistic and critical values | Break dates | Statistic and critical values | Country | Model | Estimated parameters and midpoint change date | Statistic and critical values | Break dates | Statistic and critical values |
|---|---|---|---|---|---|---|---|---|---|---|---|
| Australia | III-1 | $\hat{\lambda}_1 = 0.2673; \hat{\gamma}_1 = 152.34$<br>1930 | $\hat{S}_T = 0.0507$[b]<br>*c.v.* 10%=0.0428<br>*c.v.* 5% =0.0503<br>*c.v.* 1% =0.0684 | 1930 | $\hat{S}_T = 0.0575$[b]<br>*c.v.* 10%=0.0462<br>*c.v.* 5% =0.0539<br>*c.v.* 1% =0.0680 | Argentina | I-2 | $\hat{\lambda}_1 = 0.1358; \hat{\gamma}_1 = 167.15$<br>$\hat{\lambda}_2 = 0.3306; \hat{\gamma}_2 = 42.70$<br>1915; 1937 | $\hat{S}_T = 0.0488$[b]<br>*c.v.* 10%=0.0389<br>*c.v.* 5% =0.0449<br>*c.v.* 1% =0.0596 | 1916<br>1938 | $\hat{S}_T = 0.0488$<br>*c.v.* 10%=0.0558<br>*c.v.* 5% =0.0640<br>*c.v.* 1% =0.0817 |
| Austria | III-2 | $\hat{\lambda}_1 = 0.1549; \hat{\gamma}_1 = 132.81$<br>$\hat{\lambda}_2 = 0.1764; \hat{\gamma}_2 = 26.03$<br>1917; 1920 | $\hat{S}_T = 0.0413$[b]<br>*c.v.* 10%=0.0346<br>*c.v.* 5% =0.0404<br>*c.v.* 1% =0.0548 | 1916<br>1938 | $\hat{S}_T = 0.0410$[a]<br>*c.v.* 10% =0.0397<br>*c.v.* 5% =0.0454<br>*c.v.* 1% =0.0592 | Brazil | I-1 | $\hat{\lambda}_1 = 0.4207; \hat{\gamma}_1 = 267.96$<br>1946 | $\hat{S}_T = 0.0541$[b]<br>*c.v.* 10%=0.0410<br>*c.v.* 5% =0.0470<br>*c.v.* 1% =0.0607 | 1954 | $\hat{S}_T = 0.0834$[c]<br>*c.v.* 10%=0.0555<br>*c.v.* 5% =0.0620<br>*c.v.* 1% =0.0813 |
| Belgium | III-2 | $\hat{\lambda}_1 = 0.2123; \hat{\gamma}_1 = 58.80$<br>$\hat{\lambda}_2 = 0.4231; \hat{\gamma}_2 = 228.64$<br>1924; 1947 | $\hat{S}_T = 0.0326$[a]<br>*c.v.* 10%=0.0288<br>*c.v.* 5% =0.0338<br>*c.v.* 1% =0.0451 | 1923<br>1947 | $\hat{S}_T = 0.0361$[a]<br>*c.v.* 10%=0.0349<br>*c.v.* 5% =0.0401<br>*c.v.* 1% =0.0483 | Chile | 0 | | $\hat{S}_T = 0.0563$<br>*c.v.* 10%=0.0728<br>*c.v.* 5% =0.0873<br>*c.v.* 1% =0.1188 | | |
| Canada | 0 | | $\hat{S}_T = 0.0610$<br>*c.v.* 10%=0.0721<br>*c.v.* 5% =0.0866<br>*c.v.* 1% =0.1209 | | | China | III-2 | $\hat{\lambda}_1 = 0.0619; \hat{\gamma}_1 = 77.15$<br>$\hat{\lambda}_2 = 0.4961; \hat{\gamma}_2 = 101.23$<br>1907; 1955 | $\hat{S}_T = 0.0362$[b]<br>*c.v.* 10%=0.0301<br>*c.v.* 5% =0.0343<br>*c.v.* 1% =0.0445 | 1906<br>1957 | $\hat{S}_T = 0.0250$<br>*c.v.* 10%=0.0421<br>*c.v.* 5% =0.0485<br>*c.v.* 1% =0.0650 |
| Denmark | I-2 | $\hat{\lambda}_1 = 0.1490; \hat{\gamma}_1 = 246.82$<br>$\hat{\lambda}_2 = 0.5571; \hat{\gamma}_2 = 63.05$<br>1917; 1961 | $\hat{S}_T = 0.0509$[b]<br>*c.v.* 10%=0.0377<br>*c.v.* 5% =0.0433<br>*c.v.* 1% =0.0551 | 1917<br>1963 | $\hat{S}_T = 0.0529$[a]<br>*c.v.* 10%=0.0508<br>*c.v.* 5% =0.0575<br>*c.v.* 1% =0.0685 | Greece | III-2 | $\hat{\lambda}_1 = 0.2280; \hat{\gamma}_1 = 34.16$<br>$\hat{\lambda}_2 = 0.4329; \hat{\gamma}_2 = 174.26$<br>1925; 1948 | $\hat{S}_T = 0.0218$<br>*c.v.* 10%=0.0275<br>*c.v.* 5% =0.0318<br>*c.v.* 1% =0.0422 | 1914<br>1939 | $\hat{S}_T = 0.0455$[b]<br>*c.v.* 10%=0.0382<br>*c.v.* 5% =0.0440<br>*c.v.* 1% =0.0586 |
| Finland | III-2 | $\hat{\lambda}_1 = 0.1355; \hat{\gamma}_1 = 132.81$<br>$\hat{\lambda}_2 = 0.2730; \hat{\gamma}_2 = 26.03$<br>1915; 1930 | $\hat{S}_T = 0.0408$[b]<br>*c.v.* 10%=0.0283<br>*c.v.* 5% =0.0334<br>*c.v.* 1% =0.0443 | 1915<br>1940 | $\hat{S}_T = 0.0341$<br>*c.v.* 10%=0.0377<br>*c.v.* 5% =0.0436<br>*c.v.* 1% =0.0583 | India | I-1 | $\hat{\lambda}_1 = 0.2859; \hat{\gamma}_1 = 14.21$<br>1932 | $\hat{S}_T = 0.0615$[c]<br>*c.v.* 10%=0.0365<br>*c.v.* 5% =0.0425<br>*c.v.* 1% =0.0565 | 1931 | $\hat{S}_T = 0.0794$[a]<br>*c.v.* 10%=0.0684<br>*c.v.* 5% =0.0810<br>*c.v.* 1% =0.1098 |
| France | III-2 | $\hat{\lambda}_1 = 0.2342; \hat{\gamma}_1 = 34.16$<br>$\hat{\lambda}_2 = 0.4161; \hat{\gamma}_2 = 228.64$<br>1926; 1946 | $\hat{S}_T = 0.0411$[b]<br>*c.v.* 10%=0.0290<br>*c.v.* 5% =0.0340<br>*c.v.* 1% =0.0462 | 1922<br>1940 | $\hat{S}_T = 0.0463$[b]<br>*c.v.* 10%=0.0367<br>*c.v.* 5% =0.0413<br>*c.v.* 1% =0.0513 | Indonesia | III-2 | $\hat{\lambda}_1 = 0.4118; \hat{\gamma}_1 = 228.65$<br>$\hat{\lambda}_2 = 0.4706; \hat{\gamma}_2 = 26.03$<br>1945; 1952 | $\hat{S}_T = 0.0279$[a]<br>*c.v.* 10%=0.0259<br>*c.v.* 5% =0.0294<br>*c.v.* 1% =0.0371 | 1943<br>1947 | $\hat{S}_T = 0.0297$<br>*c.v.* 10%=0.0404<br>*c.v.* 5% =0.0460<br>*c.v.* 1% =0.0574 |
| Germany | III-2 | $\hat{\lambda}_1 = 0.4077; \hat{\gamma}_1 = 228.65$<br>$\hat{\lambda}_2 = 0.4218; \hat{\gamma}_2 = 34.15$<br>1945; 1947 | $\hat{S}_T = 0.0417$[c]<br>*c.v.* 10%=0.0270<br>*c.v.* 5% =0.0307<br>*c.v.* 1% =0.0382 | 1945<br>1951 | $\hat{S}_T = 0.0471$[b]<br>*c.v.* 10%=0.0416<br>*c.v.* 5% =0.0454<br>*c.v.* 1% =0.0544 | Mexico | III-2 | $\hat{\lambda}_1 = 0.2941; \hat{\gamma}_1 = 77.15$<br>$\hat{\lambda}_2 = 0.3137; \hat{\gamma}_2 = 44.81$<br>1933; 1935 | $\hat{S}_T = 0.0309$[a]<br>*c.v.* 10%=0.0286<br>*c.v.* 5% =0.0330<br>*c.v.* 1% =0.0440 | 1913<br>1930 | $\hat{S}_T = 0.0604$[b]<br>*c.v.* 10%=0.0474<br>*c.v.* 5% =0.0539<br>*c.v.* 1% =0.0695 |
| Italy | III-1 | $\hat{\lambda}_1 = 0.3903; \hat{\gamma}_1 = 283.53$<br>1943 | $\hat{S}_T = 0.0703$[c]<br>*c.v.* 10%=0.0406<br>*c.v.* 5% =0.0466<br>*c.v.* 1% =0.0595 | 1943 | $\hat{S}_T = 0.0712$[c]<br>*c.v.* 10%=0.0412<br>*c.v.* 5% =0.0473<br>*c.v.* 1% =0.0611 | New Zealand | III-2 | $\hat{\lambda}_1 = 0.1755; \hat{\gamma}_1 = 77.15$<br>$\hat{\lambda}_2 = 0.2793; \hat{\gamma}_2 = 132.81$<br>1920; 1931 | $\hat{S}_T = 0.0528$[b]<br>*c.v.* 10%=0.0397<br>*c.v.* 5% =0.0470<br>*c.v.* 1% =0.0637 | 1919<br>1932 | $\hat{S}_T = 0.0485$[a]<br>*c.v.* 10%=0.0439<br>*c.v.* 5% =0.0502<br>*c.v.* 1% =0.0656 |
| Japan | III-2 | $\hat{\lambda}_1 = 0.4092; \hat{\gamma}_1 = 228.64$<br>$\hat{\lambda}_2 = 0.6259; \hat{\gamma}_2 = 132.82$<br>1945; 1969 | $\hat{S}_T = 0.0424$[c]<br>*c.v.* 10%=0.0281<br>*c.v.* 5% =0.0323 | 1945<br>1969 | $\hat{S}_T = 0.0435$[b]<br>*c.v.* 10%=0.0319<br>*c.v.* 5% =0.0364 | Peru | III-2 | 1970 | $\hat{S}_T = 0.0326$[c]<br>*c.v.* 10%=0.0204<br>*c.v.* 5% =0.0229 | 1907<br>1931 | $\hat{S}_T = 0.0923$[c]<br>*c.v.* 10%=0.0432<br>*c.v.* 5% =0.0496 |
| | | | *c.v.* 1% =0.0421 | | *c.v.* 1% =0.0460 | | | | *c.v.* 1% =0.0281 | | *c.v.* 1% =0.0642 |
| Netherlands | III-2 | $\hat{\lambda}_1 = 0.1449; \hat{\gamma}_1 = 228.62$<br>$\hat{\lambda}_2 = 0.3488; \hat{\gamma}_2 = 44.82$<br>1916; 1939 | $\hat{S}_T = 0.0369^{b}$<br>*c.v.* 10%=0.0295<br>*c.v.* 5% =0.0347<br>*c.v.* 1% =0.0460 | 1943<br>1946 | $\hat{S}_T = 0.0374$<br>*c.v.* 10%=0.0432<br>*c.v.* 5% =0.0491<br>*c.v.* 1% =0.0612 | Portugal | III-2 | $\hat{\lambda}_1 = 0.1534; \hat{\gamma}_1 = 228.64$<br>$\hat{\lambda}_2 = 0.2553; \hat{\gamma}_2 = 15.12$<br>1917; 1928 | $\hat{S}_T = 0.0267^{a}$<br>*c.v.* 10%=0.0262<br>*c.v.* 5% =0.0298<br>*c.v.* 1% =0.0383 | 1917<br>1942 | $\hat{S}_T = 0.0425^{b}$<br>*c.v.* 10%=0.0372<br>*c.v.* 5% =0.0422<br>*c.v.* 1% =0.0531 |
| Spain | III-1 | $\hat{\lambda}_1 = 0.3267; \hat{\gamma}_1 = 283.53$<br>1936 | $\hat{S}_T = 0.0356$<br>*c.v.* 10%=0.0414<br>*c.v.* 5% =0.0479<br>*c.v.* 1% =0.0633 | 1936 | $\hat{S}_T = 0.0364$<br>*c.v.* 10%=0.0447<br>*c.v.* 5% =0.0522<br>*c.v.* 1% =0.0648 | Taiwan | III-2 | $\hat{\lambda}_1 = 0.2749; \hat{\gamma}_1 = 44.81$<br>$\hat{\lambda}_2 = 0.3954; \hat{\gamma}_2 = 101.22$<br>1931; 1944 | $\hat{S}_T = 0.0429^{b}$<br>*c.v.* 10%=0.0297<br>*c.v.* 5% =0.0347<br>*c.v.* 1% =0.0458 | 1919<br>1945 | $\hat{S}_T = 0.0467^{b}$<br>*c.v.* 10%=0.0359<br>*c.v.* 5% =0.0409<br>*c.v.* 1% =0.0523 |
| Sweden | III-2 | $\hat{\lambda}_1 = 0.3529; \hat{\gamma}_1 = 19.84$<br>$\hat{\lambda}_2 = 0.4118; \hat{\gamma}_2 = 28.65$<br>1939; 1945 | $\hat{S}_T = 0.0448^{c}$<br>*c.v.* 10%=0.0266<br>*c.v.* 5% =0.0306<br>*c.v.* 1% =0.0407 | 1917<br>1940 | $\hat{S}_T = 0.0650^{c}$<br>*c.v.* 10%=0.0367<br>*c.v.* 5% =0.0424<br>*c.v.* 1% =0.0525 | | | | | | |
| Switzerland | III-2 | $\hat{\lambda}_1 = 0.3725; \hat{\gamma}_1 = 19.84$<br>$\hat{\lambda}_2 = 0.4118; \hat{\gamma}_2 = 228.65$<br>1941; 1945 | $\hat{S}_T = 0.0429^{c}$<br>*c.v.* 10%=0.0265<br>*c.v.* 5% =0.0304<br>*c.v.* 1% =0.0397 | 1944<br>1946 | $\hat{S}_T = 0.0446^{a}$<br>*c.v.* 10%=0.0423<br>*c.v.* 5% =0.0481<br>*c.v.* 1% =0.0603 | | | | | | |
| United Kingdom | III-2 | $\hat{\lambda}_1 = 0.2353; \hat{\gamma}_1 = 228.65$<br>$\hat{\lambda}_2 = 0.2549; \hat{\gamma}_2 = 132.81$<br>1926; 1928 | $\hat{S}_T = 0.0437^{a}$<br>*c.v.* 10%=0.0425<br>*c.v.* 5% =0.0498<br>*c.v.* 1% =0.0681 | 1926<br>1928 | $\hat{S}_T = 0.0438$<br>*c.v.* 10%=0.0489<br>*c.v.* 5% =0.0561<br>*c.v.* 1% =0.0713 | | | | | | |
| United States | III-2 | $\hat{\lambda}_1 = 0.2812; \hat{\gamma}_1 = 132.81$<br>$\hat{\lambda}_2 = 0.4415; \hat{\gamma}_2 = 44.81$<br>1931; 1949 | $\hat{S}_T = 0.0399^{b}$<br>*c.v.* 10%=0.0251<br>*c.v.* 5% =0.0286<br>*c.v.* 1% =0.0364 | 1930<br>1949 | $\hat{S}_T = 0.0376^{b}$<br>*c.v.* 10%=0.0340<br>*c.v.* 5% =0.0372<br>*c.v.* 1% =0.0476 | | | | | | |
| **Mean Developed Countries** | III-2 | $\hat{\lambda}_1 = 0.1349; \hat{\gamma}_1 = 174.26$<br>$\hat{\lambda}_2 = 0.3734; \hat{\gamma}_2 = 58.80$<br>1915; 1941 | $\hat{S}_T = 0.0468^{b}$<br>*c.v.* 10%=0.0301<br>*c.v.* 5% =0.0353<br>*c.v.* 1% =0.0483 | 1915<br>1944 | $\hat{S}_T = 0.0440^{b}$<br>*c.v.* 10%=0.0353<br>*c.v.* 5% =0.0401<br>*c.v.* 1% =0.0518 | **Mean Developing Countries** | III-2 | $\hat{\lambda}_1 = 0.1135; \hat{\gamma}_1 = 228.65$<br>$\hat{\lambda}_2 = 0.3756; \hat{\gamma}_2 = 77.15$<br>1913; 1942 | $\hat{S}_T = 0.0451^{b}$<br>*c.v.* 10%=0.0316<br>*c.v.* 5% =0.0366<br>*c.v.* 1% =0.0487 | 1913<br>1944 | $\hat{S}_T = \hat{S}_T = 0.0437^{b}$<br>*c.v.* 10%=0.0357<br>*c.v.* 5% =0.0392<br>*c.v.* 1% =0.0492 |
| **Mean whole sample** | III-2 | $\hat{\lambda}_1 = 0.1224; \hat{\gamma}_1 = 132.81$<br>$\hat{\lambda}_2 = 0.3775; \hat{\gamma}_2 = 77.15$<br>1914; 1942 | $\hat{S}_T = 0.0481^{c}$<br>*c.v.* 10%=0.0308<br>*c.v.* 5% =0.0361<br>*c.v.* 1% =0.0480 | 1914<br>1944 | $\hat{S}_T = 0.0464^{b}$<br>*c.v.* 10%=0.0369<br>*c.v.* 5% =0.0417<br>*c.v.* 1% =0.0539 | | | | | | |

[a, b, c] denote significance at 10%, 5% and 1%, respectively.
Column "Model" indicates the model and number of changes chosen according to the sequential procedure. Model 0 (no changes); Model III (change in level and growth rate); Model I (change in level). Number of changes.
For the smooth case, $\hat{\lambda}$ and $\hat{\gamma}$ denote, respectively, the estimated relative position of the timing of the transition midpoint and the speed of transition.
*c.v.* are critical values at 10%, 5% and 1% significance levels.
The user-supplier constant, *k*, required for the data-driven device to compute the bandwidth in stationarity analysis was *k*=0.9.

**Table 2. Estimation of the number of changes in relative per capita $CO_2$ emissions**

| | JOINT ANALYSIS | | | | | | | | GROUP ANALYSIS | | | | | | | |
|---|---|---|---|---|---|---|---|---|---|---|---|---|---|---|---|---|
| | *ExpW* (Model III) | *ExpW* (Model III; unrestricted) | *ExpW*(2/1) (Model III) | *ExpW*(2/1) (Model III; unrestricted) | # Changes | *U* statistic | # Level changes $n_U$ | Model-# changes | *ExpW* (Model III) | *ExpW* (Model III; unrestricted) | *ExpW*(2/1) (Model III) | *ExpW*(2/1) (Model III; unrestricted) | # Changes | *U* statistic | # Level changes $n_U$ | Model-# changes |
| Australia | 2.2821 | 0.1480 | 2.0803 | 0.4879 | 0 | 0.4728/ 0.4878 | 0 | 0 | 2.2149 | 0.1392 | 2.0670 | 0.4790 | 0 | 0.4752/0.4645 | 0 | 0 |
| Austria | 26.8400$^{b}$ | 0.5293 | 7.9269$^{b}$ | 0.7178 | 2 | 0.5050/0.5382 | 0 | III-2 | 26.8665$^{b}$ | 0.5283 | 8.1583$^{b}$ | 0.6981 | 2 | 0.5068/0.7217$^{b}$ | 1 | III-2 |
| Belgium | 12.1880$^{b}$ | 0.2359 | 98.9810$^{b}$ | 63.3867$^{b}$ | 2 | | | III-2 | 12.2968$^{b}$ | 0.2214 | 86.3188$^{b}$ | 55.2437$^{b}$ | 2 | | | III-2 |
| Canada | 3.7410 | 0.7359 | 15.3552$^{b}$ | 13.7253$^{b}$ | 0/2 | | | 0/III-2 | 3.9293$^{a}$ | 0.7812 | 6.4537$^{a}$ | 1.5763 | 2 | 0.5660/0.5947 | 0 | III-2 |
| Denmark | 2.4178 | 1.0189 | 2.3919 | 0.4987 | 0 | 0.4322/0.4864 | 0 | 0 | 2.1186 | 0.8342 | 10.9584$^{b}$ | 9.8453$^{b}$ | 0/2 | | | 0/III-2 |
| Finland | 16.1178$^{b}$ | 0.5356 | 34.2500$^{b}$ | 7.6592$^{b}$ | 2 | | | III-2 | 16.2899$^{b}$ | 0.5289 | 33.5350$^{b}$ | 7.6978$^{b}$ | 2 | | | III-2 |
| France | 20.6453$^{b}$ | 0.2192 | 9.2510$^{b}$ | 4.6261$^{b}$ | 2 | | | III-2 | 23.4005$^{b}$ | 0.2280 | 8.8798$^{b}$ | 4.7334$^{b}$ | 2 | | | III-2 |
| Germany | 158.1051$^{b}$ | 0.0239 | 43.1962$^{b}$ | 6.4193$^{b}$ | 2 | | | III-2 | 156.3880$^{b}$ | 0.0242 | 44.8497$^{b}$ | 8.5339$^{b}$ | 2 | | | III-2 |
| Italy | 34.9502$^{b}$ | 14.0751$^{b}$ | 9.4229$^{b}$ | 6.9220$^{b}$ | 2 | | | III-2 | 35.6629$^{b}$ | 13.6332$^{b}$ | 8.4058$^{b}$ | 7.1109$^{b}$ | 2 | | | III-2 |
| Japan | 33.6801$^{b}$ | 6.7612$^{b}$ | 2.8929 | 0.6228 | 1 | | | III-1 | 35.2335$^{b}$ | 7.0527$^{b}$ | 2.8104 | 0.4614 | 1 | | | III-1 |
| Netherlands | 32.9401$^{b}$ | 0.1100 | 38.0782$^{b}$ | 2.4248 | 2 | 0.283/0.461 | 0 | III-2 | 34.0764$^{b}$ | 0.0894 | 36.9086$^{b}$ | 2.2471 | 2 | 0.2874/0.3501 | 0 | III-2 |
| Spain | 30.7319$^{b}$ | 2.4942 | 8.1162$^{b}$ | 5.7692$^{b}$ | 2 | | | III-2 | 29.9751$^{b}$ | 2.3485$^{a}$ | 8.4186$^{b}$ | 6.3604$^{b}$ | 2 | | | III-2 |
| Sweden | 28.1543$^{b}$ | 0.2119 | 33.5412$^{b}$ | 6.0201$^{b}$ | 2 | | | III-2 | 27.8401$^{b}$ | 0.2125 | 33.6731$^{b}$ | 3.8271$^{a}$ | 2 | 0.5284/0.5100 | 0 | III-2 |
| Switzerland | 37.9001$^{b}$ | 1.3750 | 21.4992 $^{b}$ | 7.256$^{b}$ | 2 | | | III-2 | 37.6518$^{b}$ | 1.3991 | 22.7205$^{b}$ | 7.3290$^{b}$ | 2 | | | III-2 |
| U.Kingdom | 41.6001$^{b}$ | 11.8509 $^{b}$ | 12.7942$^{b}$ | 5.9752$^{b}$ | 2 | | | III-2 | 24.4305$^{b}$ | 9.8135$^{b}$ | 12.9506$^{b}$ | 3.4490$^{a}$ | 2 | | | III-2 |
| U. States | 4.7742$^{a}$ | 0.4982 | 12.4582$^{b}$ | 11.7891$^{b}$ | 2 | | | III-2 | 5.6517$^{b}$ | 0.4928 | 4.7169 | 4.3072$^{b}$ | 1/2 | 0.5455/0.5821 | | III-2 |
| Argentina | 19.0893$^{b}$ | 0.4302 | 36.4289$^{b}$ | 11.3510$^{b}$ | 2 | | | III-2 | 21.9830$^{b}$ | 0.7938 | 0.3602 | 0.0179 | 1 | 0.5834/0.6856$^{b}$ | 1 | I-1 |
| Brazil | 22.0051$^{b}$ | 0.5122 | 20.1028$^{b}$ | 1.7111 | 2 | 0.416/ 0.480 | 0 | III-2 | 19.6803$^{b}$ | 0.6128 | 47.2080$^{b}$ | 37.5250$^{b}$ | 2 | | | III-2 |
| Chile | 7.8342$^{b}$ | 0.1902 | 6.1762$^{a}$ | 1.7931 | 2 | 0.3728/0.3751 | 0 | III-2 | 4.2847$^{a}$ | 0.4985 | 4.3398 | 1.2765 | 1 | 0.4439/0.4356 | | III-1 |
| China | 41.5581$^{b}$ | 39.3072$^{b}$ | 73.3222$^{b}$ | 0.6930 | 2 | | | III-2 | 49.4242$^{b}$ | 46.9100$^{b}$ | 56.8538$^{b}$ | 0.4408 | 2 | | | III-2 |
| Greece | 66.2051$^{b}$ | 0.344 | 162.781$^{b}$ | 16.507$^{b}$ | 2 | | | III-2 | 68.9341$^{b}$ | 0.3133 | 129.1426$^{b}$ | 12.6954$^{b}$ | 2 | | | III-2 |
| India | 12.4471$^{b}$ | 1.3868 | 17.8851$^{b}$ | 0.4589 | 2 | 0.4001/0.6152$^{b}$ | 1 | III-2 | 18.2073$^{b}$ | 1.3663 | 22.0431$^{b}$ | 0.1481 | 2 | 0.5398/0.6260$^{b}$ | 2 | I-2 |
| Indonesia | 22.4432$^{b}$ | 1.3502 | 19.6158$^{b}$ | 19.4111$^{b}$ | 2 | | | III-2 | 23.4436$^{b}$ | 0.9793 | 9.3610$^{b}$ | 8.2612$^{b}$ | 2 | | | III-2 |
| Mexico | 10.7579$^{b}$ | 1.1303 | 25.1418$^{b}$ | 22.0872$^{b}$ | 2 | | | III-2 | 8.4792$^{b}$ | 0.4394 | 97.0946$^{b}$ | 65.6193$^{b}$ | 2 | | | III-2 |
| New Zealand | 6.0248$^{b}$ | 0.0888 | 3.7847 | 1.4746 | 1 | 0.5452/0.3923 | 0 | III-1 | 3.9510$^{a}$ | 0.3006 | 5.3835$^{a}$ | 0.9646 | 2 | 0.5928/0.5491 | | III-2 |
| Peru | 32.0558$^{b}$ | 30.5310$^{b}$ | 11.8748$^{b}$ | 6.6388$^{b}$ | 2 | | | III-2 | 33.5335$^{b}$ | 31.7798$^{b}$ | 18.8358$^{b}$ | 9.4916$^{b}$ | 2 | | | III-2 |
| Portugal | 24.3050$^{b}$ | 0.6148 | 22.4247$^{b}$ | 20.9878$^{b}$ | 2 | | | III-2 | 18.8013$^{b}$ | 0.3185 | 31.7939$^{b}$ | 17.4203$^{b}$ | 2 | | | III-2 |
| Taiwan | 12.6888$^{b}$ | 4.3158$^{b}$ | 19.3662$^{b}$ | 0.8592 | 2 | | | III-2 | 15.3833$^{b}$ | 2.0643 | 19.1504$^{b}$ | 0.4581 | 2 | 0.7057$^{b}$/0.9333$^{b}$ | 2 | I-2 |

$\varepsilon$=0.05; $\delta$=0.5 (Perron and Yabu test) $\varepsilon$=0.1; $\delta$=0.1 (Kejriwal and Perron test).

Column *U*: *m*=0.1 (first figure)/*m*=0.15 (second figure) for the Harvey *et al.* (2010) test. *U* statistic computed at 5% level of significance. Number of level breaks in column (# Level changes, $n_U$).

$^{a,\ b}$ denote significance at 10% and 5%, respectively.

**Table 3. Results of stationarity testing on relative per capita $CO_2$ emissions. Smooth transition and break models. Joint analysis**

| | | NO CHANGE MODEL | | | | | SMOOTH TRANSITION MODEL | | | | | | | BREAK MODEL | | | | | | |
|---|---|---|---|---|---|---|---|---|---|---|---|---|---|---|---|---|---|---|---|---|
| | Model | *k*=0.5 | *k*=0.9 | *c.v.* 10% | *c.v.* 5% | *c.v.* 1% | | *k*=0.5 | *k*=0.9 | *c.v.* 10% | *c.v.* 5% | *c.v.* 1% | | | *k*=0.5 | *k*=0.9 | *c.v.* 10% | *c.v.* 5% | *c.v.* 1% | |
| Australia | 0 | 0.0459 | 0.0477 | 0.0729 | 0.0869 | 0.1180 | | | | | | | | | | | | | | |
| Austria | III-2 | | | | | | $\hat{\lambda}_1 = 0.1498; \hat{\gamma}_1 = 143.97$ <br> $\hat{\lambda}_2 = 0.1943; \hat{\gamma}_2 = 31.34$ | 0.0418[b] | 0.0418[b] | 0.0342 | 0.0403 | 0.0538 | SIC=-285.18 AIC=-314.78 Adj.$R^2$=0.81 | $\hat{\lambda}_1 = 0.1376$ <br> $\hat{\lambda}_2 = 0.1743$ | 0.0733[b] | 0.0613[a] | 0.0558 | 0.0660 | 0.0938 | SIC=-248.06 AIC=-272.28 Adj.$R^2$=0.73 |
| Belgium | III-2 | | | | | | $\hat{\lambda}_1 = 0.2089; \hat{\gamma}_1 = 44.81$ <br> $\hat{\lambda}_2 = 0.4251; \hat{\gamma}_2 = 132.78$ | 0.0216 | 0.0216 | 0.0275 | 0.0323 | 0.0424 | SIC=-497.15 AIC=-526.76 Adj.$R^2$=0.88 | $\hat{\lambda}_1 = 0.1193$ <br> $\lambda_2 = 0.2294$ | 0.0388 | 0.0390 | 0.0490 | 0.0574 | 0.0764 | SIC=-465.92 AIC=-490.15 Adj.$R^2$= 0.83 |
| Canada | 0/ III-2 | 0.0805[a] | 0.0851[a] | 0.0727 | 0.0865 | 0.1192 | $\hat{\lambda}_1 = 0.2788; \hat{\gamma}_1 = 23.63$ <br> $\hat{\lambda}_2 = 0.4263; \hat{\gamma}_2 = 55.13$ | 0.0234[a] | 0.0234[a] | 0.0233 | 0.0267 | 0.0347 | SIC=-566.80 AIC=-596.41 Adj.$R^2$=0.88 | $\hat{\lambda}_1 = 0.1468$ <br> $\lambda_2 = 0.3578$ | 0.0525[b] | 0.0525[b] | 0.0389 | 0.0437 | 0.0578 | SIC=-541.76 AIC=-565.99 Adj.$R^2$=0.83 |
| Denmark | 0 | 0.0550 | 0.0544 | 0.0721 | 0.0865 | 0.1202 | | | | | | | | | | | | | | |
| Finland | III-2 | | | | | | $\hat{\lambda}_1 = 0.1361; \hat{\gamma}_1 = 132.82$ <br> $\hat{\lambda}_2 = 0.2474; \hat{\gamma}_2 = 19.84$ | 0.0477[c] | 0.0477[c] | 0.0262 | 0.0303 | 0.0399 | SIC=-251.53 AIC=-281.13 Adj.$R^2$=0.96 | $\hat{\lambda}_1 = 0.1284$ <br> $\lambda_2 = 0.1927$ | 0.0661[b] | 0.0661[b] | 0.0544 | 0.0610 | 0.0813 | SIC=-226.88 AIC=-251.10 Adj.$R^2$=0.95 |
| France | III-2 | | | | | | $\hat{\lambda}_1 = 0.2549; \hat{\gamma}_1 = 19.84$ <br> $\hat{\lambda}_2 = 0.4111; \hat{\gamma}_2 = 228.65$ | 0.0323[a] | 0.0323[a] | 0.0285 | 0.0332 | 0.0442 | SIC=-541.19 AIC=-570.80 Adj.$R^2$= 0.88 | $\hat{\lambda}_1 = 0.1743$ <br> $\lambda_2 = 0.3578$ | 0.0532[b] | 0.0521[b] | 0.0381 | 0.0434 | 0.0554 | SIC=-468.05 AIC=-492.27 Adj.$R^2$=0.83 |
| Germany | III-2 | | | | | | $\hat{\lambda}_1 = 0.3960; \hat{\gamma}_1 = 77.36$ <br> $\hat{\lambda}_2 = 0.4227; \hat{\gamma}_2 = 121.54$ | 0.0558[c] | 0.0558[c] | 0.0324 | 0.0372 | 0.0480 | SIC=-454.20 AIC=-483.81 Adj.$R^2$=0.90 | $\hat{\lambda}_1 = 0.4018$ <br> $\lambda_2 = 0.4771$ | 0.0374 | 0.0368 | 0.0396 | 0.0452 | 0.0599 | SIC=-494.04 AIC=-518.26 Adj.$R^2$=0.90 |
| Italy | III-2 | | | | | | $\hat{\lambda}_1 = 0.3812; \hat{\gamma}_1 = 132.82$ <br> $\hat{\lambda}_2 = 0.4110; \hat{\gamma}_2 = 228.65$ | 0.0569[c] | 0.0569[c] | 0.0369 | 0.0426 | 0.0545 | SIC=-405.75 AIC=-435.35 Adj.$R^2$=0.98 | $\hat{\lambda}_1 = 0.3853$ <br> $\lambda_2 = 0.4128$ | 0.0662[c] | 0.0620[c] | 0.0410 | 0.0450 | 0.0583 | SIC=-403.04 AIC=-427.26 Adj.$R^2$=0.97 |
| Japan | III-1 | | | | | | $\hat{\lambda} = 0.4143; \gamma = 283.53$ | 0.0482[b] | 0.0488[b] | 0.0409 | 0.0470 | 0.0605 | SIC=-474.33 AIC=-493.17 Adj.$R^2$=0.97 | $\hat{\lambda}_1 = 0.4128$ | 0.0449[a] | 0.0460[a] | 0.0449 | 0.0522 | 0.0649 | SIC=-471.79 AIC=-490.63 Adj.$R^2$=0.97 |
| Netherlands | III-2 | | | | | | $\hat{\lambda}_1 = 0.3119; \hat{\gamma}_1 = 19.84$ <br> $\hat{\lambda}_2 = 0.4203; \hat{\gamma}_2 = 174.26$ | 0.0219 | 0.0219 | 0.0276 | 0.0319 | 0.0432 | SIC=-425.17 AIC=-454.77 Adj.$R^2$=0.65 | $\hat{\lambda}_1 = 0.2752$ <br> $\lambda_2 = 0.4128$ | 0.0358[a] | 0.0358[a] | 0.0356 | 0.0394 | 0.0502 | SIC=-415.46 AIC=-439.68 Adj.$R^2$=0.60 |
| Spain | III-2 | | | | | | $\hat{\lambda}_1 = 0.3275; \hat{\gamma}_1 = 28.64$ <br> $\hat{\lambda}_2 = 0.3344; \hat{\gamma}_2 = 174.26$ | 0.0479[b] | 0.0454[b] | 0.0387 | 0.0445 | 0.0583 | SIC=-450.12 AIC=-479.73 Adj.$R^2$= 0.97 | $\hat{\lambda}_1 = 0.3211$ <br> $\lambda_2 = 0.3395$ | 0.0387 | 0.0388 | 0.0408 | 0.0467 | 0.0635 | SIC=-443.28 AIC=-467.50 Adj.$R^2$=0.97 |
| Sweden | III-2 | | | | | | $\hat{\lambda}_1 = 0.3513; \hat{\gamma}_1 = 19.84$ <br> $\hat{\lambda}_2 = 0.4183; \hat{\gamma}_2 = 228.64$ | 0.0489[c] | 0.0489[c] | 0.0269 | 0.0311 | 0.0408 | SIC=-370.90 AIC=-400.50 Adj.$R^2$= 0.80 | $\hat{\lambda}_1 = 0.1468$ <br> $\lambda_2 = 0.3578$ | 0.0731[c] | 0.0728[c] | 0.0363 | 0.0412 | 0.0528 | SIC=-348.01 AIC=-372.23 Adj. $R^2$=0.74 |
| Switzerland | III-2 | | | | | | $\hat{\lambda}_1 = 0.3236; \hat{\gamma}_1 = 30.99$ <br> $\hat{\lambda}_2 = 0.4197; \hat{\gamma}_2 = 288.08$ | 0.0514[c] | 0.0514[c] | 0.0298 | 0.0347 | 0.0463 | SIC=-433.60 AIC=-463.21 Adj.$R^2$= 0.79 | $\hat{\lambda}_1 = 0.3486$ <br> $\lambda_2 = 0.4128$ | 0.0554[b] | 0.0497[b] | 0.0399 | 0.0446 | 0.0568 | SIC=-405.62 AIC=-429.84 Adj.$R^2$=0.76 |
| United Kingdom | III-2 | | | | | | $\hat{\lambda}_1 = 0.2739; \hat{\gamma}_1 = 121.54$ <br> $\hat{\lambda}_2 = 0.6396; \hat{\gamma}_2 = 31.34$ | 0.0236[b] | 0.0236[b] | 0.0193 | 0.0214 | 0.0261 | SIC=-532.94 AIC=-562.55 Adj.$R^2$= 0.98 | $\hat{\lambda}_1 = 0.2385$ <br> $\lambda_2 = 0.6972$ | 0.0271 | 0.0271 | 0.0318 | 0.0349 | 0.0414 | SIC=-536.40 AIC=-560.63 Adj.$R^2$=0.97 |
| United States | III-2 | | | | | | $\hat{\lambda}_1 = 0.2923; \hat{\gamma}_1 = 29.62$ <br> $\hat{\lambda}_2 = 0.4221; \hat{\gamma}_2 = 213.78$ | 0.0330[a] | 0.0330[a] | 0.0287 | 0.0335 | 0.04474 | SIC=-591.93 AIC=-621.53 Adj.$R^2$= 0.96 | $\hat{\lambda}_1 = 0.1376$ <br> $\lambda_2 = 0.3670$ | 0.0491[b] | 0.0491[b] | 0.0376 | 0.0441 | 0.0548 | SIC=-573.64 AIC=-597.86 Adj.$R^2$=0.95 |
| Argentina | III-2 | | | | | | $\hat{\lambda}_1 = 0.1500; \hat{\gamma}_1 = 96.96$ <br> $\hat{\lambda}_2 = 0.1892; \hat{\gamma}_2 = 52.10$ | 0.0456[a] | 0.0434[a] | 0.0405 | 0.0483 | 0.0649 | SIC=-391.26 AIC=-420.86 Adj.$R^2$=0.82 | $\hat{\lambda}_1 = 0.1193$ <br> $\lambda_2 = 0.1651$ | 0.0421 | 0.0404 | 0.0568 | 0.0670 | 0.0670 | SIC=-382.54 AIC=-406.76 Adj.$R^2$=0.79 |
| Brazil | III-2 | | | | | | $\hat{\lambda}_1 = 0.1407; \hat{\gamma}_1 = 136.07$ <br> $\hat{\lambda}_2 = 0.2572; \hat{\gamma}_2 = 39.28$ | 0.0368[a] | 0.0373[a] | 0.0337 | 0.0396 | 0.0540 | SIC=-470.18 <br> AIC=-499.78 <br> Adj.$R^2$= 0.97 | $\hat{\lambda}_1 = 0.1193$ <br> $\lambda_2 = 0.4037$ | 0.0315 | 0.0351 | 0.0370 | 0.0424 | 0.0519 | SIC=-472.26 <br> AIC=-496.48 <br> Adj.$R^2$ =0.96 |
| Chile | III-2 | | | | | | $\hat{\lambda}_1 = 0.2128; \hat{\gamma}_1 = 54.66$ <br> $\hat{\lambda}_2 = 0.7218; \hat{\gamma}_2 = 39.52$ | 0.0283[c] | 0.0283[c] | 0.0188 | 0.0210 | 0.0262 | SIC=-442.45 <br> AIC=-472.06 <br> Adj.$R^2$=0.74 | $\hat{\lambda}_1 = 0.2477$ <br> $\lambda_2 = 0.7982$ | 0.0320 | 0.0320 | 0.0326 | 0.0359 | 0.0425 | SIC=-447.55 <br> AIC=-471.77 <br> Adj.$R^2$= 0.73 |
| China | III-2 | | | | | | $\hat{\lambda}_1 = 0.0568; \hat{\gamma}_1 = 86.61$ <br> $\hat{\lambda}_2 = 0.5074; \hat{\gamma}_2 = 121.53$ | 0.0416[b] | 0.0416[b] | 0.0321 | 0.0371 | 0.0481 | SIC=-192.58 <br> AIC=-222.18 <br> Adj.$R^2$=0.95 | $\hat{\lambda}_1 = 0.0458$ <br> $\lambda_2 = 0.5229$ | 0.0336 | 0.0336 | 0.0404 | 0.0466 | 0.0611 | SIC=-202.31 <br> AIC=-226.53 <br> Adj.$R^2$=0.96 |
| Greece | III-2 | | | | | | $\hat{\lambda}_1 = 0.2321; \hat{\gamma}_1 = 29.62$ <br> $\hat{\lambda}_2 = 0.4323; \hat{\gamma}_2 = 143.97$ | 0.0208 | 0.0208 | 0.0271 | 0.0314 | 0.0413 | SIC=-78.30 <br> AIC=-107.92 <br> Adj.$R^2$= 0.89 | $\hat{\lambda}_1 = 0.1193$ <br> $\lambda_2 = 0.3486$ | 0.0475[b] | 0.0457[a] | 0.0400 | 0.0464 | 0.0594 | SIC=-91.42 <br> AIC=-115.65 <br> Adj.$R^2$=0.90 |
| India | III-2 | | | | | | $\hat{\lambda}_1 = 0.2219; \hat{\gamma}_1 = 49.23$ <br> $\hat{\lambda}_2 = 0.6377; \hat{\gamma}_2 = 117.86$ | 0.0322[c] | 0.0320[b] | 0.0223 | 0.0251 | 0.0321 | SIC=-532.71 <br> AIC=-562.31 <br> Adj.$R^2$= 0.98 | $\hat{\lambda}_1 = 0.2018$ <br> $\lambda_2 = 0.6330$ | 0.0408[b] | 0.0408[b] | 0.0325 | 0.0356 | 0.0435 | SIC=-522.50 <br> AIC=-546.73 <br> Adj.$R^2$= 0.98 |
| Indonesia | III-2 | | | | | | $\hat{\lambda}_1 = 0.4172; \hat{\gamma}_1 = 121.53$ <br> $\hat{\lambda}_2 = 0.4700; \hat{\gamma}_2 = 25.03$ | 0.0331[c] | 0.0331[b] | 0.0235 | 0.0264 | 0.0327 | SIC=-322.71 <br> AIC=-352.32 <br> Adj.$R^2$=0.94 | $\hat{\lambda}_1 = 0.4037$ <br> $\lambda_2 = 0.5597$ | 0.0292 | 0.0292 | 0.0353 | 0.0396 | 0.0483 | SIC=-305.93 <br> AIC=-330.15 <br> Adj.$R^2$ =0.92 |
| Mexico | III-2 | | | | | | $\hat{\lambda}_1 = 0.1171; \hat{\gamma}_1 = 283.55$ <br> $\hat{\lambda}_2 = 0.1257; \hat{\gamma}_2 = 43.98$ | 0.0395 | 0.0395 | 0.0460 | 0.0540 | 0.0747 | SIC=-310.49 <br> AIC=-340.09 <br> Adj.$R^2$=0.93 | $\hat{\lambda}_1 = 0.1100$ <br> $\lambda_2 = 0.1925$ | 0.0373 | 0.0374 | 0.0539 | 0.0622 | 0.0855 | SIC=-324.49 <br> AIC=-348.71 <br> Adj.$R^2$=0.93 |
| New Zealand | III-1 | | | | | | $\hat{\lambda} = 0.3861; \hat{\gamma} = 77.36$ | 0.0473[b] | 0.0457[b] | 0.0359 | 0.0408 | 0.0532 | SIC=-455.77 <br> AIC=-474.61 <br> Adj.$R^2$= 0.71 | $\hat{\lambda} = 0.4495$ | 0.0442[a] | 0.0443[a] | 0.0436 | 0.0510 | 0.0601 | SIC=-451.90 <br> AIC=-470.74 <br> Adj.$R^2$=0.70 |
| Peru | III-2 | | | | | | $\hat{\lambda}_1 = 0.0001; \hat{\gamma}_1 = 58.33$ <br> $\hat{\lambda}_2 = 0.4358; \hat{\gamma}_2 = 21.11$ | 0.0399[c] | 0.0399[c] | 0.0224 | 0.0252 | 0.0317 | SIC=-304.39 <br> AIC=-333.99 <br> Adj.$R^2$=0.86 | $\hat{\lambda}_1 = 0.0642$ <br> $\lambda_2 = 0.3945$ | 0.0670[c] | 0.0670[c] | 0.0393 | 0.0454 | 0.0568 | SIC=-286.06 <br> AIC=-310.29 <br> Adj.$R^2$=0.83 |
| Portugal | III-2 | | | | | | $\hat{\lambda}_1 = 0.1534; \hat{\gamma}_1 = 283.52$ <br> $\hat{\lambda}_2 = 0.2761; \hat{\gamma}_2 = 12.70$ | 0.0406[c] | 0.0406[c] | 0.0247 | 0.0281 | 0.0358 | SIC=-437.56 <br> AIC=-467.16 <br> Adj.$R^2$=0.96 | $\hat{\lambda}_1 = 0.1468$ <br> $\lambda_2 = 0.3578$ | 0.0625[c] | 0.0625[c] | 0.0379 | 0.0438 | 0.0539 | SIC=-426.64 <br> AIC=-450.86 <br> Adj.$R^2$=0.95 |
| Taiwan | III-2 | | | | | | $\hat{\lambda}_1 = 0.1530; \hat{\gamma}_1 = 96.96$ <br> $\hat{\lambda}_2 = 0.4054; \hat{\gamma}_2 = 283.52$ | 0.0557[c] | 0.0515[c] | 0.0319 | 0.0371 | 0.0489 | SIC=-428.31 <br> AIC=-457.91 <br> Adj.$R^2$=0.99 | $\hat{\lambda}_1 = 0.1560$ <br> $\lambda_2 = 0.4037$ | 0.0566[c] | 0.0544[c] | 0.0356 | 0.0401 | 0.0498 | SIC=-406.21 <br> AIC=-430.44 <br> Adj.$R^2$=0.98 |

[a, b, c] denote significance at 10%, 5% and 1%, respectively.

**Table 4. Results of stationarity testing on relative per capita $CO_2$ emissions. Smooth transition and break models. Group analysis**

| | | NO CHANGE MODEL | | | | | SMOOTH TRANSITION MODEL | | | | | | | BREAK MODEL | | | | | | |
|---|---|---|---|---|---|---|---|---|---|---|---|---|---|---|---|---|---|---|---|---|
| | Model | *k*=0.5 | *k*=0.9 | *c.v.* 10% | *c.v.* 5% | *c.v.* 1% | | *k*=0.5 | *k*=0.9 | *c.v.* 10% | *c.v.* 5% | *c.v.* 1% | | | *k*=0.5 | *k*=0.9 | *c.v.* 10% | *c.v.* 5% | *c.v.* 1% | |
| Australia | 0 | 0.0459 | 0.0477 | 0.0734 | 0.0873 | 0.1210 | | | | | | | | | | | | | | |
| Austria | III-2 | | | | | | $\hat{\lambda}_1 = 0.1498; \hat{\gamma}_1 = 143.97$<br>$\hat{\lambda}_2 = 0.1943; \hat{\gamma}_2 = 31.34$ | 0.0436[b] | 0.0433[b] | 0.0340 | 0.0399 | 0.0544 | SIC=-285.73<br>AIC=-315.33<br>Adj.$R^2$=0.82 | $\hat{\lambda}_1 = 0.1376$<br>$\lambda_2 = 0.1743$ | 0.0720[b] | 0.0603[b] | 0.0538 | 0.0633 | 0.0856 | SIC=-248.06<br>AIC=-272.28<br>Adj.$R^2$=0.73 |
| Belgium | III-2 | | | | | | $\hat{\lambda}_1 = 0.2090; \hat{\gamma}_1 = 43.98$<br>$\hat{\lambda}_2 = 0.4221; \hat{\gamma}_2 = 283.53$ | 0.0209 | 0.0209 | 0.0291 | 0.0342 | 0.0453 | SIC=-502.23<br>AIC=-531.83<br>Adj.$R^2$=0.89 | $\hat{\lambda}_1 = 0.1192$<br>$\lambda_2 = 0.2294$ | 0.0392 | 0.0391 | 0.0489 | 0.0563 | 0.0770 | SIC=-465.93<br>AIC=-490.15<br>Adj.$R^2$=0.83 |
| Canada | III-2 | | | | | | $\hat{\lambda}_1 = 0.2765; \hat{\gamma}_1 = 23.63$<br>$\hat{\lambda}_2 = 0.4242; \hat{\gamma}_2 = 52.10$ | 0.0235[a] | 0.0235[a] | 0.0230 | 0.0267 | 0.0349 | SIC= -563.81<br>AIC=-593.41<br>Adj.$R^2$=0.82 | $\hat{\lambda}_1 = 0.1468$<br>$\lambda_2 = 0.3578$ | 0.0542[b] | 0.0542[b] | 0.0372 | 0.0424 | 0.0561 | SIC=-545.63<br>AIC=-569.85<br>Adj.$R^2$=0.77 |
| Denmark | 0/<br>III-2 | 0.0492 | 0.0492 | 0.0730 | 0.0874 | 0.1226 | $\hat{\lambda}_1 = 0.2609; \hat{\gamma}_1 = 69.10$<br>$\hat{\lambda}_2 = 0.5596; \hat{\gamma}_2 = 143.97$ | 0.0221 | 0.0221 | 0.0230 | 0.0259 | 0.0321 | SIC= -500.95<br>AIC=-530.55<br>Adj.$R^2$=0.87 | $\hat{\lambda}_1 = 0.2569$<br>$\lambda_2 = 0.5596$ | 0.0221 | 0.0221 | 0.0317 | 0.0347 | 0.0423 | SIC= -503.46<br>AIC=-527.69<br>Adj.$R^2$=0.86 |
| Finland | III-2 | | | | | | $\hat{\lambda}_1 = 0.1397; \hat{\gamma}_1 = 96.96$<br>$\hat{\lambda}_2 = 0.2197; \hat{\gamma}_2 = 18.85$ | 0.0463[c] | 0.0463[c] | 0.0268 | 0.0310 | 0.0406 | SIC=-248.36<br>AIC=-277.96<br>Adj.$R^2$=0.96 | $\hat{\lambda}_1 = 0.1284$<br>$\lambda_2 = 0.1927$ | 0.0645[b] | 0.0645[b] | 0.0532 | 0.0637 | 0.0794 | SIC=-226.88<br>AIC=-251.10<br>Adj.$R^2$=0.95 |
| France | III-2 | | | | | | $\hat{\lambda}_1 = 0.2434; \hat{\gamma}_1 = 21.11$<br>$\hat{\lambda}_2 = 0.4108; \hat{\gamma}_2 = 283.53$ | 0.0322[a] | 0.0322[a] | 0.0291 | 0.0343 | 0.0457 | SIC=-544.16<br>AIC=-573.77<br>Adj.$R^2$=0.88 | $\hat{\lambda}_1 = 0.3394$<br>$\lambda_2 = 0.4037$ | 0.0439[a] | 0.0416[a] | 0.0398 | 0.0443 | 0.0543 | SIC=-471.46<br>AIC=-495.68<br>Adj.$R^2$=0.75 |
| Germany | III-2 | | | | | | $\hat{\lambda}_1 = 0.3953; \hat{\gamma}_1 = 77.36$<br>$\hat{\lambda}_2 = 0.4236; \hat{\gamma}_2 = 121.54$ | 0.0598 | 0.0598 | 0.0323 | 0.0373 | 0.0479 | SIC=-452.96<br>AIC=-482.56<br>Adj.$R^2$=0.87 | $\hat{\lambda}_1 = 0.4036$<br>$\lambda_2 = 0.4770$ | 0.03739 | 0.0368 | 0.0390 | 0.0449 | 0.0588 | SIC=-494.04<br>AIC=-518.26<br>Adj.$R^2$=0.90 |
| Italy | III-2 | | | | | | $\hat{\lambda}_1 = 0.3789; \hat{\gamma}_1 = 121.54$<br>$\hat{\lambda}_2 = 0.4121; \hat{\gamma}_2 = 283.53$ | 0.0579[c] | 0.0579[c] | 0.0368 | 0.0420 | 0.0544 | SIC=-404.08<br>AIC=-433.68<br>Adj.$R^2$=0.97 | $\hat{\lambda}_1 = 0.3853$<br>$\lambda_2 = 0.4128$ | 0.0654[c] | 0.0625[c] | 0.0433 | 0.0474 | 0.0575 | SIC=-403.04<br>AIC=-427.26<br>Adj.$R^2$=0.97 |
| Japan | III-1 | | | | | | $\hat{\lambda}_1 = 0.4143; \hat{\gamma}_1 = 283.53$ | 0.0482[b] | 0.0488[b] | 0.0410 | 0.0471 | 0.0613 | SIC=-474.33<br>AIC=-493.17<br>Adj.$R^2$=0.97 | $\hat{\lambda}_1 = 0.4128$ | 0.0437[a] | 0.0452[a] | 0.0431 | 0.0489 | 0.0637 | SIC=-471.79<br>AIC=-490.63<br>Adj.$R^2$=0.97 |
| Netherlands | III-2 | | | | | | $\hat{\lambda}_1 = 0.2704; \hat{\gamma}_1 = 26.45$<br>$\hat{\lambda}_2 = 0.4205; \hat{\gamma}_2 = 283.53$ | 0.0192 | 0.0192 | 0.0288 | 0.0338 | 0.0458 | SIC=-429.37<br>AIC=-458.97<br>Adj.$R^2$=0.67 | $\hat{\lambda}_1 = 0.2752$<br>$\lambda_2 = 0.4128$ | 0.0334 | 0.0334 | 0.0355 | 0.0401 | 0.0498 | SIC=-415.46<br>AIC=-439.68<br>Adj.$R^2$=0.60 |
| Spain | III-2 | | | | | | $\hat{\lambda}_1 = 0.3196; \hat{\gamma}_1 = 108.56$<br>$\hat{\lambda}_2 = 0.3434; \hat{\gamma}_2 = 283.53$ | 0.0405[a] | 0.0404[a] | 0.0370 | 0.0427 | 0.0568 | SIC=-383.01<br>AIC=-412.62<br>Adj.$R^2$=0.94 | $\hat{\lambda}_1 = 0.3211$<br>$\lambda_2 = 0.3395$ | 0.0375 | 0.0379 | 0.0434 | 0.0498 | 0.0665 | SIC=-443.28<br>AIC=-467.50<br>Adj.$R^2$=0.97 |
| Sweden | III-2 | | | | | | $\hat{\lambda}_1 = 0.3150; \hat{\gamma}_1 = 31.34$<br>$\hat{\lambda}_2 = 0.4203; \hat{\gamma}_2 = 283.53$ | 0.0534[c] | 0.0580[c] | 0.0300 | 0.0350 | 0.0469 | SIC=-379.61<br>AIC=-409.21<br>Adj.$R^2$=0.82 | $\hat{\lambda}_1 = 0.1468$<br>$\lambda_2 = 0.3578$ | 0.0735[c] | 0.0726[c] | 0.0381 | 0.0436 | 0.0545 | SIC=-348.01<br>AIC=-372.23<br>Adj. $R^2$=0.74 |
| Switzerland | III-2 | | | | | | $\hat{\lambda}_1 = 0.3403; \hat{\gamma}_1 = 25.01$<br>$\hat{\lambda}_2 = 0.4195; \hat{\gamma}_2 = 283.53$ | 0.0534[c] | 0.0534[c] | 0.0292 | 0.0338 | 0.0449 | SIC=-435.04<br>AIC=-464.64<br>Adj.$R^2$=0.83 | $\hat{\lambda}_1 = 0.3486$<br>$\lambda_2 = 0.4128$ | 0.0546[c] | 0.0491[b] | 0.0398 | 0.0439 | 0.0539 | SIC=-405.62<br>AIC=-429.84<br>Adj.$R^2$=0.76 |
| United Kingdom | III-2 | | | | | | $\hat{\lambda}_1 = 0.2744; \hat{\gamma}_1 = 136.07$<br>$\hat{\lambda}_2 = 0.6331; \hat{\gamma}_2 = 33.16$ | 0.0228[b] | 0.0228[b] | 0.0198 | 0.0221 | 0.0270 | SIC=-534.41<br>AIC=-564.02<br>Adj.$R^2$=0.97 | $\hat{\lambda}_1 = 0.2294$<br>$\lambda_2 = 0.6972$ | 0.0375[b] | 0.0375[b] | 0.0310 | 0.0338 | 0.0405 | SIC=-537.66<br>AIC=-561.88<br>Adj.$R^2$=0.96 |
| United States | III-2 | | | | | | $\hat{\lambda}_1 = 0.1449; \hat{\gamma}_1 = 226.20$<br>$\hat{\lambda}_2 = 0.3704; \hat{\gamma}_2 = 96.96$ | 0.0700[c] | 0.0700[c] | 0.0318 | 0.0375 | 0.0501 | SIC=-594.64<br>AIC=-624.24<br>Adj.$R^2$=0.94 | $\hat{\lambda}_1 = 0.1376$<br>$\lambda_2 = 0.3578$ | 0.0430[b] | 0.0430[b] | 0.0371 | 0.0429 | 0.0548 | SIC=-567.71<br>AIC=-591.93<br>Adj.$R^2$=0.92 |
| Argentina | I-1 | | | | | | $\hat{\lambda}_1 = 0.1366; \hat{\gamma}_1 = 283.53$ | 0.0532 | 0.0552 | 0.0592 | 0.0696 | 0.0945 | SIC=-360.19<br>AIC=-376.33<br>Adj.$R^2$=0.67 | $\hat{\lambda}_1 = 0.1284$ | 0.0490 | 0.0494 | 0.0614 | 0.0703 | 0.0912 | SIC=-355.83<br>AIC=-371.98<br>Adj.$R^2$=0.66 |
| Brazil | III-2 | | | | | | $\hat{\lambda}_1 = 0.1429; \hat{\gamma}_1 = 152.34$<br>$\hat{\lambda}_2 = 0.2346; \hat{\gamma}_2 = 33.16$ | 0.0422[b] | 0.0420[b] | 0.0331 | 0.0389 | 0.0509 | SIC=-456.14<br>AIC=-485.75<br>Adj.$R^2$=0.83 | $\hat{\lambda}_1 = 0.1927$<br>$\hat{\lambda}_2 = 0.4128$ | 0.0367 | 0.0370[a] | 0.0349 | 0.0382 | 0.0466 | SIC=-452.83<br>AIC=-477.05<br>Adj.$R^2$=0.82 |
| Chile | III-1 | | | | | | $\hat{\lambda}_1 = 0.7101; \hat{\gamma}_1 = 43.98$ | 0.0284 | 0.0317 | 0.0344 | 0.0404 | 0.0530 | SIC=-479.76<br>AIC=-498.60<br>Adj.$R^2$=0.89 | $\hat{\lambda}_1 = 0.7431$ | 0.0594[b] | 0.0605[b] | 0.0467 | 0.0531 | 0.0707 | SIC=-448.78<br>AIC=-467.62<br>Adj.$R^2$=0.86 |
| China | III-2 | | | | | | $\hat{\lambda}_1 = 0.0656; \hat{\gamma}_1 = 73.11$<br>$\hat{\lambda}_2 = 0.4094; \hat{\gamma}_2 = 283.53$ | 0.0526[c] | 0.0526[c] | 0.0333 | 0.0383 | 0.0501 | SIC=-205.43<br>AIC=-235.03<br>Adj.$R^2$=0.94 | $\hat{\lambda}_1 = 0.0184$<br>$\hat{\lambda}_2 = 0.0459$ | 0.0402 | 0.0562 | 0.0686 | 0.0803 | 0.1060 | SIC=-217.17<br>AIC=-241.39<br>Adj.$R^2$=0.94 |
| Greece | III-2 | | | | | | $\hat{\lambda}_1 = 0.2271; \hat{\gamma}_1 = 31.34$<br>$\hat{\lambda}_2 = 0.4318; \hat{\gamma}_2 = 190.95$ | 0.0187 | 0.0187 | 0.0277 | 0.0321 | 0.0425 | SIC=-77.05<br>AIC=-106.66<br>Adj.$R^2$=0.85 | $\hat{\lambda}_1 = 0.1193$<br>$\hat{\lambda}_2 = 0.3486$ | 0.0475[b] | 0.0457[b] | 0.0382 | 0.0444 | 0.0604 | SIC=-92.76<br>AIC=-116.98<br>Adj.$R^2$=0.86 |
| India | I-2 | | | | | | $\hat{\lambda}_1 = 0.6955; \hat{\gamma}_1 = 49.24$<br>$\hat{\lambda}_2 = 0.8563; \hat{\gamma}_2 = 52.10$ | 0.0288 | 0.0313 | 0.0344 | 0.04010 | 0.0530 | SIC=-479.21<br>AIC=-503.43<br>Adj.$R^2$=0.90 | $\hat{\lambda}_1 = 0.1101$<br>$\hat{\lambda}_2 = 0.4495$ | 0.0662[b] | 0.0579[a] | 0.0529 | 0.0597 | 0.0840 | SIC=-477.07<br>AIC=-495.91<br>Adj.$R^2$=0.81 |
| Indonesia | III-2 | | | | | | $\hat{\lambda}_1 = 0.4165; \hat{\gamma}_1 = 102.59$<br>$\hat{\lambda}_2 = 0.4688; \hat{\gamma}_2 = 26.45$ | 0.0229[a] | 0.0229[a] | 0.0224 | 0.0254 | 0.0316 | SIC=-319.19<br>AIC=-348.79<br>Adj.$R^2$=0.83 | $\hat{\lambda}_1 = 0.3853$<br>$\hat{\lambda}_2 = 0.4220$ | 0.0327 | 0.0327 | 0.0416 | 0.0467 | 0.0562 | SIC=-323.81<br>AIC=-348.03<br>Adj.$R^2$=0.82 |
| Mexico | III-2 | | | | | | $\hat{\lambda}_1 = 0.1197; \hat{\gamma}_1 = 283.53$<br>$\hat{\lambda}_2 = 0.1337; \hat{\gamma}_2 = 55.12$ | 0.0420 | 0.0420 | 0.0468 | 0.0557 | 0.0776 | SIC=-322.26<br>AIC=-351.87<br>Adj.$R^2$=0.88 | $\hat{\lambda}_1 = 0.1100$<br>$\hat{\lambda}_2 = 0.1925$ | 0.0372 | 0.0372 | 0.0527 | 0.0625 | 0.0774 | SIC=-314.23<br>AIC=-338.45<br>Adj.$R^2$=0.86 |
| New Zealand | III-2 | | | | | | $\hat{\lambda}_1 = 0.3745; \hat{\gamma}_1 = 91.64$<br>$\hat{\lambda}_2 = 0.7668; \hat{\gamma}_2 = 180.46$ | 0.0227 | 0.0297[b] | 0.0249 | 0.0281 | 0.0361 | SIC=-577.21<br>AIC=-606.81<br>Adj.$R^2$=0.98 | $\hat{\lambda}_1 = 0.3761$<br>$\hat{\lambda}_2 = 0.7706$ | 0.0337[a] | 0.0369[b] | 0.0321 | 0.0356 | 0.0446 | SIC=-561.49<br>AIC=-585.71<br>Adj.$R^2$=0.98 |
| Peru | III-2 | | | | | | $\hat{\lambda}_1 = 0.0001; \hat{\gamma}_1 = 34.99$<br>$\hat{\lambda}_2 = 0.5631; \hat{\gamma}_2 = 5.21$ | 0.0353[c] | 0.0353[c] | 0.0223 | 0.0251 | 0.0315 | SIC=-304.52<br>AIC=-334.12<br>Adj.$R^2$=0.81 | $\hat{\lambda}_1 = 0.0642$<br>$\hat{\lambda}_2 = 0.5963$ | 0.0428[a] | 0.0428[a] | 0.0389 | 0.0443 | 0.0550 | SIC=-286.45<br>AIC=-310.67<br>Adj.$R^2$=0.76 |
| Portugal | III-2 | | | | | | $\hat{\lambda}_1 = 0.1528; \hat{\gamma}_1 = 283.53$<br>$\hat{\lambda}_2 = 0.2972; \hat{\gamma}_2 = 13.43$ | 0.0306[b] | 0.0306[b] | 0.0237 | 0.0268 | 0.0338 | SIC=-420.99<br>AIC=-450.59<br>Adj.$R^2$=0.80 | $\hat{\lambda}_1 = 0.1468$<br>$\hat{\lambda}_2 = 0.3578$ | 0.0545[b] | 0.0544[b] | 0.0392 | 0.0451 | 0.0541 | SIC=-403.85<br>AIC=-428.07<br>Adj.$R^2$=0.75 |
| Taiwan | I-2 | | | | | | $\hat{\lambda}_1 = 0.1288; \hat{\gamma}_1 = 46.48$<br>$\hat{\lambda}_2 = 0.4027; \hat{\gamma}_2 = 288.08$ | 0.0839[c] | 0.0772[c] | 0.0435 | 0.0514 | 0.0699 | SIC=-403.06<br>AIC=-427.28<br>Adj.$R^2$=0.97 | $\hat{\lambda}_1 = 0.1376$<br>$\hat{\lambda}_2 = 0.4037$ | 0.0771[b] | 0.0715[b] | 0.0555 | 0.0649 | 0.0922 | SIC=-337.23<br>AIC=-356.07<br>Adj.$R^2$=0.93 |

[a, b, c] denote significance at 10%, 5% and 1%, respectively

**Table 5. Change dates, parameter estimates, and convergence/divergence classification. Joint analysis**

| | SMOOTH TRANSTION MODEL | | | | | | | | | | | BREAK MODEL | | | | | | | | | | |
|---|---|---|---|---|---|---|---|---|---|---|---|---|---|---|---|---|---|---|---|---|---|---|
| | Dates (Mid-point) | Parameter estimates | | | | | | | Convergence/Divergence | | | Dates | Parameter estimates | | | | | | | Convergence/Divergence | | |
| | | $\hat{\delta}_1$ | $\hat{\eta}_1$ | $\hat{\delta}_2$ | $\hat{\eta}_2$ | $\hat{\delta}_3$ | $\hat{\eta}_3$ | $\hat{\beta}$ | Reg. 1 | Reg. 2 | Reg. 3 | | $\hat{\delta}_1$ | $\hat{\eta}_1$ | $\hat{\delta}_2$ | $\hat{\eta}_2$ | $\hat{\delta}_3$ | $\hat{\eta}_3$ | $\hat{\beta}$ | Reg. 1 | Reg. 2 | Reg. 3 |
| Australia | | 0.265<br>(6.829)<br>+ | 0.735<br>(4.151)<br>+ | | | | | -0.142<br>(-0.820) | *D* | | | | | | | | | | | | | |
| Austria | 1916<br>1920 | | | | | | | | | | | 1915<br>1919 | | | | | | | | | | |
| Belgium | 1923<br>1947 | 1.203<br>(32.987)<br>+ | -3.797<br>(-5.808)<br>- | 1.575<br>(22.844)<br>+ | -5.809<br>(-8.480)<br>- | 0.827<br>(16.798)<br>+ | -3.526<br>(-5.545)<br>- | 1.459<br>(3.387) | *C* | *C* | *C* | 1914<br>1926 | 1.212<br>(36.083)<br>+ | -2.912<br>(-4.382)<br>- | 0.445<br>(12.203)<br>+ | 7.381<br>(14.330)<br>+ | 1.185<br>(21.466)<br>+ | -1.375<br>(-4.107)<br>- | 0.095<br>(0.390) | *C* | *D* | *C* |
| Canada | 1931<br>1947 | 0.669<br>(24.989)<br>+ | 4.065<br>(8.604)<br>+ | -0.005<br>(-0.051)<br>+ | 11.524<br>(11.014)<br>+ | 0.942<br>(30.783)<br>+ | -0.208<br>(-0.436)<br>- | -0.103<br>(-0.335) | *D* | *d* | *c* | 1917<br>1940 | | | | | | | | | | |
| Denmark | | -0.217<br>(-4.109)<br>- | 2.396<br>(10.190)<br>+ | | | | | -2.059<br>(-9.475) | *C* | | | | | | | | | | | | | |
| Finland | 1916<br>1946 | | | | | | | | | | | 1915<br>1944 | | | | | | | | | | |
| France | 1929<br>1946 | 0.275<br>(10.406)<br>+ | -2.635<br>(-3.698)<br>+ | 1.672<br>(9.312)<br>+ | -11.134<br>(-7.457)<br>- | 0.652<br>(11.001)<br>+ | 0.502<br>(0.632)<br>- | -1.332<br>(-2.445) | *D* | *C* | *c* | 1943<br>1945 | | | | | | | | | | |
| Germany | 1945<br>1947 | | | | | | | | | | | 1944<br>1953 | 0.957<br>(33.214)<br>+ | -0.751<br>(-2.452)<br>+ | -0.615<br>(-4.185)<br>- | 8.769<br>(6.793)<br>+ | 0.121<br>(0.887)<br>+ | -3.814<br>(-3.599)<br>- | 2.146<br>(2.959) | *D* | *C* | *c* |
| Italy | 1944<br>1949 | | | | | | | | | | | 1942<br>1945 | | | | | | | | | | |
| Japan | 1946 | | | | | | | | | | | 1945 | -1.632<br>(-29.573)<br>- | 5.623<br>(15.165)<br>+ | -0.148<br>(-1.531)<br>- | 13.356<br>(11.324)<br>+ | | | -7.860<br>(-9.413) | *C* | *c* | |
| Netherlands | 1934<br>1947 | 0.173<br>(6.150)<br>+ | -0.667<br>(-0.876)<br>+ | 1.363<br>(4.887)<br>+ | -12.977<br>(-4.171)<br>- | 0.632<br>(9.367)<br>+ | 2.248<br>(2.428)<br>- | -1.689<br>(-2.718) | *d* | *C* | *C* | 1930<br>1945 | 0.071<br>(1.132)<br>+ | 1.340<br>(2.835)<br>+ | 0.906<br>(11.774)<br>+ | -4.878<br>(-4.903)<br>- | 0.618<br>(12.530)<br>+ | 2.862<br>(4.572)<br>- | -2.091<br>(-4.749) | *D* | *C* | *C* |
| Spain | 1937<br>1942 | | | | | | | | | | | 1936<br>1938 | -1.304<br>(-23.678)<br>- | 0.233<br>(0.518)<br>- | -3.645<br>(-44.415)<br>- | 109.165<br>(204.431)<br>+ | -1.259<br>(-21.510)<br>- | 4.005<br>(3.609)<br>+ | -1.323<br>(-1.659) | *d* | *C* | *C* |
| Sweden | 1942<br>1947 | | | | | | | | | | | 1944<br>1947 | | | | | | | | | | |
| Switzerland | 1942<br>1947 | | | | | | | | | | | 1944<br>1947 | | | | | | | | | | |
| United Kingdom | 1928 | | | | | | | | | | | 1926 | 1.406 | -0.040 | 1.460 | 4.253 | 3.384 | 9.673 | -6.452 | *c* | *C* | *C* |
| | 1946 | | | | | | | | | | | 1976 | (23.718)<br>+ | (-0.096)<br>- | (74.680)<br>+ | (7.659)<br>- | (12.405)<br>+ | (9.599)<br>- | (-10.821) | | | |
| United States | 1932<br>1947 | 1.455<br>(54.375)<br>+ | 1.018<br>(3.793)<br>- | 0.668<br>(8.522)<br>+ | 7.445<br>(7.926)<br>+ | 1.215<br>(37.168)<br>+ | -3.561<br>(-9.158)<br>- | 1.731<br>(6.445) | *C* | *C* | *C* | 1937<br>1945 | | | | | | | | | | |
| | | | | | | | | | | | | | | | | | | | | | | |
| Argentina | 1917<br>1921 | -1.363<br>(-16.430)<br>- | 4.548<br>(4.204)<br>- | -3.637<br>(-10.858)<br>- | 65.819<br>(5.298)<br>+ | -1.325<br>(-14.750)<br>- | 3.666<br>(5.206)<br>+ | -2.447<br>(-4.197) | *D* | *C* | *C* | 1914<br>1918 | -1.442<br>(-18.431)<br>- | 5.262<br>(5.114)<br>+ | -0.806<br>(-9.358)<br>- | -41.529<br>(-9.716)<br>- | -1.334<br>(-17.041)<br>- | 3.052<br>(4.911)<br>+ | -2.020<br>(-3.750) | *C* | *D* | *C* |
| Brazil | 1916<br>1929 | -2.862<br>(-108.630)<br>- | 2.188<br>(3.543)<br>- | -3.227<br>(-40.628)<br>- | 11.373<br>(7.521)<br>+ | -2.836<br>(-47.508)<br>- | 6.203<br>(9.576)<br>+ | -3.268<br>(-6.904) | *D* | *C* | *C* | 1914<br>1945 | -2.910<br>(-109.334)<br>- | 2.648<br>(6.061)<br>+ | -2.908<br>(-32.416)<br>- | 2.435<br>(4.473)<br>+ | -1.981<br>(-27.330)<br>- | 4.657<br>(6.259)<br>+ | -2.260<br>(-4.228) | *C* | *C* | *C* |
| Chile | 1924<br>1979 | | | | | | | | | | | 1928<br>1988 | -1.392<br>(-30.851)<br>- | 2.401<br>(6.217)<br>+ | -1.037<br>(-26.794)<br>- | 6.688<br>(9.785)<br>- | -2.973<br>(7.033)<br>- | 14.147<br>(11.394)<br>+ | -6.450<br>(-9.700) | *C* | *D* | *C* |
| China | 1902<br>1946 | | | | | | | | | | | 1905<br>1957 | -10.082<br>(-29.367)<br>- | 99.218<br>(8.426)<br>+ | -4.292<br>(-25.936)<br>- | 1.027<br>(0.656)<br>+ | -3.197<br>(-6.441)<br>- | -3.695<br>(-0.889)<br>+ | 4.574<br>(1.791) | *C* | *c* | *c* |
| Greece | 1926<br>1948 | -3.103<br>(-10.486)<br>- | -12.406<br>(-2.810)<br>+ | 1.681<br>(2.468)<br>- | -36.569<br>(-7.048)<br>- | -0.967<br>(-3.998)<br>- | 10.852<br>(3.157)<br>+ | -4.957<br>(-2.186) | *C* | *D* | *C* | 1913<br>1938 | | | | | | | | | | |
| India | 1919<br>1970 | | | | | | | | | | | 1919<br>1970 | | | | | | | | | | |
| Indonesia | 1903<br>1946 | | | | | | | | | | | 1945<br>1962 | -4.122<br>(-35.639)<br>- | 5.073<br>(5.638)<br>+ | -3.336<br>(-10.527)<br>- | 18.271<br>(4.070)<br>+ | -1.479<br>(-2.337)<br>- | 14.377<br>(3.957)<br>+ | -6.400<br>(-2.787) | *C* | *C* | *C* |
| Mexico | 1913<br>1914 | -3.133<br>(-42.260)<br>- | -1.778<br>(-0.856)<br>+ | -4.672<br>(-15.839)<br>- | 126.002<br>(4.765)<br>- | -2.264<br>(-11.773)<br>- | 4.570<br>(4.442)<br>+ | -2.402<br>(-3.173) | *c* | *D* | *C* | 1912<br>1921 | -3.419<br>(-31.726)<br>- | 11.995<br>(7.658)<br>+ | -3.458<br>(-39.959)<br>- | 28.598<br>(18.668)<br>+ | -1.902<br>(-13.873)<br>- | 3.818<br>(3.775)<br>+ | -1.850<br>(-2.499) | *C* | *C* | *C* |
| New Zealand | 1942 | | | | | | | | | | | 1950 | 0.608<br>(10.749)<br>+ | -3.475<br>(-8.064)<br>- | -0.776<br>(-4.870)<br>- | -6.933<br>(-5.408)<br>+ | | | 4.808<br>(5.247) | *C* | *C* | |
| Peru | 1903<br>1937 | | | | | | | | | | | 1932<br>1934 | | | | | | | | | | |
| Portugal | 1917<br>1940 | | | | | | | | | | | 1917<br>1940 | | | | | | | | | | |
| Taiwan | 1918<br>1945 | | | | | | | | | | | 1944<br>1949 | | | | | | | | | | |

Parameter estimates and *t*-statistic (between parentheses). Symbols "+" and "-" indicate the sign of $\hat{\delta}_k$ or that of the value of the estimated function at the beginning of the regime (for the break and smooth transition models respectively), and the sign of the median derivative of the estimated function (column $\hat{\eta}_k$ ).

**Table 6. Change dates, parameter estimates, and convergence/divergence classification. Group analysis**

| | SMOOTH TRANSTION MODEL | | | | | | | | | | | BREAK MODEL | | | | | | | | | | |
|---|---|---|---|---|---|---|---|---|---|---|---|---|---|---|---|---|---|---|---|---|---|---|
| | Dates (Mid-point) | Parameter estimates | | | | | | | Convergence/Divergence | | | Dates | Parameter estimates | | | | | | | Convergence/Divergence | | |
| | | $\hat{\delta}_1$ | $\hat{\eta}_1$ | $\hat{\delta}_2$ | $\hat{\eta}_2$ | $\hat{\delta}_3$ | $\hat{\eta}_3$ | $\hat{\beta}$ | Reg. 1 | Reg. 2 | Reg. 3 | | $\hat{\delta}_1$ | $\hat{\eta}_1$ | $\hat{\delta}_2$ | $\hat{\eta}_2$ | $\hat{\delta}_3$ | $\hat{\eta}_3$ | $\hat{\beta}$ | Reg. 1 | Reg. 2 | Reg. 3 |
| Australia | | -0.183<br>(-4.304)<br>- | 0.573<br>(2.961)<br>+ | | | | | 0.223<br>(1.202) | *C* | | | | | | | | | | | | | |
| Austria | 1917<br>1938 | | | | | | | | | | | 1915<br>1919 | | | | | | | | | | |
| Belgium | 1923<br>1947 | 0.744<br>(-19.593)<br>+ | -3.863<br>(-6.018)<br>- | 1.103<br>(16.810)<br>+ | -5.998<br>(-9.416)<br>- | 0.313<br>(7.620)<br>+ | -3.591<br>(-7.420)<br>- | 1.780<br>(5.331) | *C* | *C* | *C* | 1914<br>1926 | 0.748<br>(-21.038)<br>+ | -2.894<br>(-4.163)<br>- | -0.017<br>(-0.478)<br>- | 7.188<br>(14.807)<br>+ | 0.704<br>(12.861)<br>+ | -1.630<br>(-4.839)<br>- | 0.535<br>(2.123) | *C* | *c* | *C* |
| Canada | 1931<br>1947 | 0.206<br>(7.872)<br>+ | 4.054<br>(8.393)<br>+ | -0.545<br>(-5.462)<br>+ | 11.203<br>(10.513)<br>+ | 0.392<br>(13.376)<br>+ | -0.757<br>(-1.528)<br>+ | 0.534<br>(1.672) | *D* | *D* | *d* | 1918<br>1945 | | | | | | | | | | |
| Denmark | 1929<br>1961 | -0.568<br>(-15.556)<br>- | 1.008<br>(3.054)<br>+ | 0.004<br>(0.089)<br>- | 0.266<br>(0.439)<br>- | 0.355<br>(1.748)<br>+ | 0.102<br>(0.092)<br>- | -0.536<br>(-0.762) | *C* | *d* | *c* | 1929<br>1961 | -0.603<br>(-13.979)<br>- | 1.359<br>(4.004)<br>+ | -0.012<br>(-0.284)<br>- | 0.861<br>(1.557)<br>+ | 0.464<br>(2.574)<br>+ | 0.883<br>(0.853)<br>- | -1.006<br>(-1.531) | *C* | *c* | *c* |
| Finland | 1915<br>1945 | | | | | | | | | | | 1914<br>1945 | | | | | | | | | | |
| France | 1927<br>1945 | -0.169<br>(-6.743)<br>- | -2.556<br>(-4.183)<br>+ | 1.047<br>(7.793)<br>+ | -9.602<br>(-8.238)<br>- | 0.128<br>(2.426)<br>- | 0.346<br>(0.523)<br>- | -0.955<br>(-2.080) | *C* | *C* | *D* | 1938<br>1945 | -0.372<br>(-4.354)<br>- | 1.540<br>(3.915)<br>+ | -0.104<br>(-0.883)<br>- | -5.284<br>(-1.847)<br>- | 0.102<br>(1.807)<br>+ | 0.318<br>(0.436)<br>- | -0.904<br>(-1.757) | *C* | *d* | *C* |
| Germany | 1944<br>1947 | 0.511<br>(16.941)<br>+ | -0.864<br>(-3.059)<br>- | -0.257<br>(-0.841)<br>+ | -72.393<br>(-4.974)<br>- | -0.140<br>(-1.872)<br>- | -2.835<br>(-2.986)<br>- | 1.779<br>(2.777) | *C* | *C* | *D* | 1944<br>1953 | 0.511<br>(17.294)<br>+ | -0.951<br>(-3.080)<br>+ | -1.164<br>(-8.329)<br>- | 8.750<br>(7.358)<br>+ | -0.437<br>(-3.163)<br>- | -4.207<br>(-3.975)<br>- | 2.674<br>(3.683) | *D* | *C* | *D* |
| Italy | 1942<br>1946 | | | | | | | | | | | 1942<br>1945 | | | | | | | | | | |
| Japan | 1946 | | | | | | | | | | | 1945 | -2.089<br>(-39.104)<br>- | 5.638<br>(15.896)<br>+ | -0.509<br>(-5.287)<br>- | 14.094<br>(13.732)<br>+ | | | 8.081<br>(-10.953) | *C* | *C* | |
| Netherlands | 1930<br>1946 | -0.254<br>(-9.565)<br>- | -0.824<br>(-1.317)<br>+ | 0.748<br>(4.834)<br>+ | -8.108<br>(-5.620)<br>- | 0.121<br>(2.264)<br>- | 2.077<br>(3.260)<br>+ | -1.323<br>(-3.028) | *c* | *C* | *C* | 1930<br>1945 | -0.384<br>(-6.516)<br>- | 1.212<br>(2.683)<br>+ | 0.406<br>(5.178)<br>+ | -5.112<br>(-5.344)<br>- | 0.088<br>(1.882)<br>+ | 2.506<br>(4.032)<br>+ | -1.583<br>(-3.641) | *C* | *C* | *D* |
| Spain | 1935<br>1938 | -1.819<br>(-36.240)<br>- | 0.618<br>(1.552)<br>+ | -3.187<br>(-9.089)<br>- | 0.078<br>(0.003)<br>- | -1.763<br>(-30.748<br>- | 2.850<br>(2.822)<br>+ | -0.245<br>(-0.334) | *c* | *d* | *C* | 1935<br>1938 | -1.808<br>(-41.988)<br>- | 0.509<br>(1.597)<br>+ | -3.158<br>(-12.820)<br>- | 26.525<br>(2.052)<br>+ | -1.777<br>(-32.287)<br>- | 3.398<br>(3.264)<br>+ | -0.624<br>(-0.836) | *c* | *C* | *C* |
| Sweden | 1916<br>1946 | | | | | | | | | | | 1916<br>1945 | | | | | | | | | | |
| Switzerland | 1937<br>1946 | | | | | | | | | | | 1944<br>1946 | | | | | | | | | | |
| United Kingdom | 1926 | | | | | | | | | | | 1926 | | | | | | | | | | |
| | 1977 | | | | | | | | | | | 1977 | | | | | | | | | | |
| United States | 1917<br>1945 | | | | | | | | | | | 1937<br>1945 | | | | | | | | | | |
| | | | | | | | | | | | | | | | | | | | | | | |
| Argentina | 1915 | 0.222<br>(2.571)<br>+ | 3.379<br>(2.890)<br>+ | 0.023<br>(0.128)<br>+ | 4.308<br>(3.805)<br>- | | | -4.207<br>(-4.680) | *D* | *c* | | 1915 | 0.173<br>(1.957)<br>+ | 3.712<br>(3.353)<br>+ | 0.003<br>(0.016)<br>+ | 4.217<br>(4.113)<br>- | | | -4.137<br>(-5.041) | *D* | *C* | |
| Brazil | 1920<br>1946 | | | | | | | | | | | 1922<br>1945 | -1.227<br>(-25.470)<br>- | -0.381<br>(-0.553)<br>- | -1.005<br>(-29.740)<br>- | 1.099<br>(2.293)<br>- | -0.435<br>(-5.502)<br>- | 3.626<br>(3.478)<br>- | -2.765<br>(-3.841) | *d* | *D* | *D* |
| Chile | 1978 | 0.188<br>(3.985)<br>+ | 1.135<br>(3.887)<br>+ | 0.246<br>(1.096)<br>- | 4.340<br>(6.140)<br>+ | | | -1.606<br>(-3.796) | *D* | *c* | | 1919 | | | | | | | | | | |
| China | 1902<br>1946 | | | | | | | | | | | 1903<br>1906 | -9.499<br>(-18.924)<br>- | 158.370<br>(6.102)<br>+ | -5.897<br>(-20.214)<br>- | 106.647<br>(7.076)<br>+ | -3.067<br>(-20.723)<br>- | 4.968<br>(5.013)<br>+ | -1.746<br>(-1.861) | *C* | *C* | *C* |
| Greece | 1925<br>1948 | -1.548<br>(-5.351)<br>- | -12.440<br>(-2.782)<br>- | 3.133<br>(4.863)<br>- | -32.703<br>(-6.758)<br>- | 0.727<br>(3.265)<br>- | 12.763<br>(4.115)<br>+ | -7.387<br>(-3.603) | *C* | *D* | *C* | 1913<br>1938 | | | | | | | | | | |
| India | 1976<br>1994 | -2.131<br>(-30.389)<br>- | 3.228<br>(8.145)<br>+ | 0.305<br>(1.028)<br>- | 8.499<br>(11.620)<br>+ | 1.540<br>(3.666)<br>- | 9.503<br>(8.384)<br>+ | -4.074<br>(-7.382) | *C* | *c* | *C* | 1919<br>1949 | | | | | | | | | | |
| Indonesia | 1946<br>1952 | -2.593<br>(-25.488)<br>- | 5.702<br>(6.598)<br>+ | -2.375<br>(-4.544)<br>- | 71.715<br>(6.751)<br>+ | -1.270<br>(-4.586)<br>- | 16.786<br>(4.874)<br>+ | -8.766<br>(-4.083) | *C* | *C* | *C* | 1942<br>1946 | -2.493<br>(-23.979)<br>- | 3.297<br>(4.642)<br>+ | 0.007<br>(0.030)<br>+ | -90.501<br>(-38.480)<br>- | -2.053<br>(-13.676)<br>- | 3.852<br>(1.727)<br>+ | -1.244<br>(-0.851) | *C* | *c* | *C* |
| Mexico | 1914<br>1915 | -1.401<br>(-9.480)<br>- | -2.293<br>(-0.585)<br>+ | -3.712<br>(-6.533)<br>- | 145.823<br>(3.493)<br>+ | -0.594<br>(-3.382)<br>- | 4.506<br>(4.577)<br>+ | -3.597<br>(-4.889) | *c* | *C* | *C* | 1912<br>1921 | -1.822<br>(-16.833)<br>- | 10.663<br>(6.929)<br>+ | -2.286<br>(-13.285)<br>- | 33.950<br>(11.475)<br>+ | -0.283<br>(-2.337)<br>- | 3.797<br>(4.362)<br>+ | -3.067<br>(-4.783) | *C* | *C* | *C* |
| New Zealand | 1920<br>1978 | | | | | | | | | | | 1913<br>1919 | | | | | | | | | | |
| Peru | 1903<br>1935 | | | | | | | | | | | 1908<br>1965 | -2.846<br>(-11.726)<br>- | 22.670<br>(4.806)<br>+ | -0.827<br>(-7.611)<br>- | -2.963<br>(-2.388)<br>+ | -2.330<br>(-4.008)<br>- | -12.859<br>(-4.437)<br>- | 6.231<br>(3.407) | *C* | *C* | *D* |
| Portugal | 1918<br>1950 | | | | | | | | | | | 1916<br>1918 | | | | | | | | | | |
| Taiwan | 1917<br>1945 | | | | | | | | | | | 1944<br>1949 | | | | | | | | | | |

Parameter estimates and *t*-statistic (between parentheses). Symbols "+" and "-" indicate the sign of $\hat{\delta}_k$ or that of the value of the estimated function at the beginning of the regime (for the break and smooth transition models respectively), and the sign of the median derivative of the estimated function (column $\hat{\eta}_k$ ).